\documentclass[longauth]{aa}  
\usepackage{graphicx}
\usepackage{txfonts}
\usepackage{xcolor}
\usepackage{booktabs}
\usepackage[font=small]{caption}
\newcommand{\kms}{\,km\,s$^{-1}$}
\newcommand{\cii}{[C~II]}

\begin{document}

   \title{The ALPINE--ALMA \cii\ survey: the luminosity function of serendipitous \cii\ line emitters at $z\sim 5$}


   \author{Federica Loiacono\inst{1,2,10,11}\fnmsep\thanks{\email{federica.loiacono2@unibo.it}}, Roberto Decarli\inst{2}, Carlotta Gruppioni\inst{2}, Margherita Talia\inst{1,2}, Andrea Cimatti\inst{2,3}, Gianni Zamorani\inst{2}, Francesca Pozzi\inst{1}, Lin Yan\inst{4}, Brian C. Lemaux\inst{20}, Dominik A. Riechers\inst{5,6}, Olivier Le F$\grave{\rm e}$vre\inst{19}, Matthieu B$\acute{\rm e}$thermin\inst{19}, Peter Capak\inst{7}, Paolo Cassata\inst{14,21}, Andreas Faisst\inst{7}, Daniel Schaerer\inst{15}, John D. Silverman\inst{22,23}, Sandro Bardelli\inst{2}, M$\acute{\rm e}$d$\acute{\rm e}$ric Boquien\inst{18}, Sandra Burkutean\inst{17}, Miroslava Dessauges-Zavadsky\inst{15}, Yoshinobu Fudamoto\inst{15}, Seiji Fujimoto\inst{8,9}, Michele Ginolfi\inst{15}, Nimish P. Hathi\inst{12}, Gareth C. Jones\inst{10,11}, Yana Khusanova\inst{19}, Anton M. Koekemoer\inst{12}, Guilaine Lagache\inst{19}, Marcella Massardi\inst{17}, Pascal Oesch\inst{15}, Michael Romano\inst{14,21}, Livia Vallini\inst{13}, Daniela Vergani\inst{2} and Elena Zucca\inst{2}
          }

\institute{Università di Bologna, Dipartimento di Fisica e Astronomia (DIFA), 
   via Gobetti 93/2, I-40129 Bologna, Italy\\
\and INAF -- Osservatorio di Astrofisica e Scienza dello Spazio, via Gobetti 93/3, I-40129, Bologna, Italy\\
\and INAF -- Osservatorio  Astrofisico  di  Arcetri,  Largo  E.  Fermi  5,  I-50125, Firenze, Italy\\
\and The Caltech Optical Observatories, California Institute of Technology, Pasadena, CA 91125, USA\\
\and Department of Astronomy, Cornell University, Space Sciences Building, Ithaca, NY 14853, USA\\
\and Max-Planck-Institut f\"ur Astronomie, K\"onigstuhl 17, D-69117 Heidelberg, Germany\\
\and IPAC,  California  Institute  of  Technology,  1200  East  CaliforniaBoulevard, Pasadena, CA 91125, USA\\
\and The Cosmic Dawn Center, University of Copenhagen, Vibenshuset, Lyngbyvej 2, DK-2100 Copenhagen, Denmark\\
\and Niels Bohr Institute, University of Copenhagen, Lyngbyvej 2, DK-2100 Copenhagen, Denmark\\
\and Cavendish Laboratory, University of Cambridge, 19 J. J. Thomson Ave., Cambridge CB3 0HE, UK\\
\and Kavli Institute for Cosmology, University of Cambridge, Madingley Road, Cambridge CB3 0HA, UK\\
\and Space Telescope Science Institute, 3700 San Martin Dr., 
Baltimore, MD 21218, USA\\
\and Leiden Observatory, Leiden University, PO Box 9500, 2300 RA Leiden, The Netherlands\\
\and Dipartimento  di  Fisica  e  Astronomia,  Università  di  Padova, Vicolo dell’Osservatorio 3, I-35122, Padova, Italy\\
\and Observatoire de Gen$\grave{\rm e}$ve, Universit$\acute{\rm e}$ de Gen$\grave{\rm e}$ve, 51 Ch. des Maillettes, 1290 Versoix, Switzerland\\
\and INAF, Istituto di Radioastronomia, via Piero Gobetti 101, I-40129 Bologna, Italy\\
\and Centro de Astronomia (CITEVA), Universidad de Antofagasta, Avenida Angamos 601, Antofagasta, Chile\\
\and Aix Marseille Univ, CNRS, CNES, LAM, Marseille, France
\and Department of Physics, University of California, Davis, One Shields Ave., Davis, CA 95616, USA
\and INAF, Osservatorio Astronomico di Padova, vicolo dell'Osservatorio 5, I-35122 Padova, Italy
\and Kavli Institute for the Physics and Mathematics of the Universe, The University of Tokyo Kashiwa, Chiba 277-8583, Japan
\and Department of Astronomy, School of Science, The University of Tokyo, 7-3-1 Hongo, Bunkyo, Tokyo 113-0033, Japan
}

   \date{submitted to A \&\ A}

 
  \abstract
{We present the first \cii\ 158 $\mu$m luminosity function (LF) at $z\sim 5$ from a sample of serendipitous lines detected in the ALMA Large Program to INvestigate \cii\ at Early times (ALPINE). 
A search performed over the 118 ALPINE pointings revealed several serendipitous lines. Based on their fidelity, we selected 14 lines for the final catalog. According to the redshift of their counterparts, we identified eight out of 14 detections as \cii\ lines at $z\sim 5$, and two as CO transitions at lower redshifts. The remaining four lines have an elusive identification in the available catalogs and we considered them as \cii\ candidates. We used the eight confirmed \cii\ and the four \cii\ candidates to build one of the first \cii\ LFs at $z\sim 5$.
We found that 11 out of these 12 sources have a redshift very similar to that of the ALPINE target in the same pointing, suggesting the presence of overdensities around the targets. Therefore, we split the sample in two (a $``$clustered$"$ and $``$field$"$ sub--sample) according to their redshift separation and built two separate LFs.
Our estimates suggest that there could be an evolution of the \cii\ LF between $z \sim 5$ and $z \sim 0$.
By converting the \cii\ luminosity to star formation rate we evaluated the cosmic star formation rate density (SFRD) at $z\sim 5$.
The clustered sample results in a SFRD $\sim 10$ times higher than previous measurements from UV--selected galaxies. On the other hand, from the field sample (likely representing the average galaxy population) we derived a SFRD $\sim 1.6$ higher compared to current estimates from UV surveys but compatible within the errors. Because of the large uncertainties, observations of larger samples are necessary to better constrain the SFRD at $z\sim 5$.
This study represents one of the first efforts aimed at characterizing the demography of \cii\ emitters at $z\sim 5$ using a mm--selection of galaxies.}
   \keywords{galaxies: evolution --
                galaxies: high-redshift -- galaxies: ISM -- submillimeter: galaxies -- galaxies: luminosity function, mass function}
\titlerunning{The \cii\ luminosity function at $z\sim 5$ from ALPINE serendipitous lines
}
\authorrunning{Loiacono et al.
}
\maketitle
\section{Introduction}
Our quest on the early phases of galaxy evolution cannot prescind from the study of cold gas.
High--redshift galaxies are indeed more gas--rich than present day objects with gas fractions up to unity, as witnessed by large observing campaigns (e.g. \citealt{2018ApJ...853..179T}).
The rate at which the Universe form stars varies significantly across cosmic time \citep{2014ARA&A..52..415M}; however, the drivers of this trend are still poorly known. 
Up to $z\sim 3$, we have a robust understanding of the star formation history, thanks to more than twenty years of multi--wavelength investigations (e.g., \citealt{2003ApJ...587L..89T, 2005ApJ...619L..47S, 2012A&A...539A..31C, 2013MNRAS.432...23G, 2013A&A...553A.132M, 2015ApJ...803...34B, 2016PASA...33...37F, 2018ApJ...855..105O, 2020MNRAS.493.2059B}). Nevertheless at $z>3$ our constraints are almost exclusively based on observations sampling the rest--frame ultraviolet (UV) emission, which is very sensitive to dust reddening. Studies at longer wavelengths (e.g., \citealt{2013MNRAS.432....2K, 2016MNRAS.461.1100R, 2017A&A...602A...5N, 2018A&A...614A..39M}) hint to the presence of a population of gas-- and dust--rich galaxies that may be missed by the UV selection. However, the demography of such dusty galaxies, and therefore their role in shaping the cosmic star formation rate density at $z>3$, is still very uncertain.\\
\indent Over the last few years we have been witnessing a true revolution, with the Atacama Large Millimeter/submillimeter Array (ALMA) opening a window on the high--$z$ obscured Universe. Thanks to its unprecedented sensitivity, ALMA allows us to detect for the first time the dust continuum and the bright infrared (IR) lines in normal galaxies at $z > 3$ and constrain the cosmic star formation history (\citealt{2016ApJ...833...72B, 2017ApJ...837..150S, 2019ApJ...887..235L}). In particular, the \cii\ 158 $\mu$m line can easily be detected because it is one of brightest galaxy lines in the IR, radiating up to a hundredth of the entire far--infrared luminosity of a galaxy \citep{2013ApJ...774...68D}, and it is conveniently redshifted into atmospheric relatively transparent windows. This line is mainly excited by collision with neutral hydrogen atoms in the so--called photo--dissociation regions (PDRs; \citealt{1999RvMP...71..173H}) and in the neutral diffuse gas \citep{2003ApJ...587..278W}. Nevertheless, it can also trace diffuse ionized gas where it is excited by collisions with free electrons (e.g. \citealt{2012A&A...548A..20C}). 
Thanks to its brightness, \cii\ is a powerful tool to derive accurate redshifts of distant galaxies (e.g. \citealt{2012Natur.486..233W, 2013Natur.496..329R, 2015Natur.522..455C}). Spatially resolved observations of this line can be used to characterize the kinematics of the cold interstellar medium (ISM; \citealt{2018Natur.553..178S, 2019MNRAS.487.3007K}).
Besides, when other lines are also available, flux ratios can be used to study the physical properties of the ISM in terms of gas density, strength of the radiation field and excitation source (e.g., \citealt{2016ApJ...832..151P, 2018ApJ...861...43P, 2019ApJ...881...63N}).
Finally, \cii\ has also been found to be a star--formation rate (SFR) indicator at low and possibly at high--$z$ by observations (\citealt{2014A&A...568A..62D, 2014ApJ...796...63M, 2018MNRAS.478.1170C, 2019ApJ...881..124M, 2020arXiv200200979S}) and models predictions \citep{2015ApJ...813...36V, 2018A&A...609A.130L}.\\ 
\indent The main limitation of mm interferometers such as ALMA is their relatively small field of view, which makes surveys of blank fields expensive in terms of telescope time. Most of the studies at high--$z$ have
therefore focused on the exploration of properties of $``$targeted$"$ galaxies that were pre--selected based on their stellar mass, SFR and/or IR luminosity (e.g. \citet{2015A&A...577A..46D, 2018ApJ...853..179T}). 
These kind of studies have been instrumental to shape our understanding of the connection between the inner gas reservoirs and the build--up of galaxies. However, the pre--selection may introduce biases associated to our prior knowledge of the emitting systems.
On the other hand, $``$blind$"$ surveys, as well as serendipitous discoveries in observations targeting other sources, aid to circumvent selection biases, thus enabling a proper census of the cold gas properties in a volume-limited region of the universe \citep{2016ApJ...833...69D, 2019ApJ...872....7R}.
In particular, blind selections of lines in the mm--domain are sensitive to heavily obscured galaxies that can be missed in the UV surveys. Properly accounting for these objects is crucial when estimating global quantities such as the cosmic SFRD and building luminosity functions.\\
\indent Recently, the ALMA Large Program to INvestigate \cii\ at Early times (ALPINE) has been completed \citep{2019arXiv191009517L, 2020arXiv200200962B, 2019arXiv191201621F}.
This project aims at studying the \cii\ emission in 118 spectroscopically confirmed and UV--selected star--forming galaxies at $4 < z < 6$.
A search for spectral lines in the 118 ALPINE pointings unveiled a wealth of unexpected lines, i.e. serendipitous discoveries in a wide redshift range. Most of the lines are due to \cii\ emission. We use these lines to build the \cii\ luminosity function at $z \sim 5$. This is the first \cii\ LF based on galaxies purely selected for their \cii\ emission. On the other hand, the companion paper of Yan et al. (2020) presents the \cii\ LF from the UV--selected central targets.
Despite being well constrained at $z \sim 0$ from statistical samples, at high--redshift the number density of \cii\ emitters represents an uncharted territory. A knowledge of their LF is crucial to constrain the semi--analytical models and cosmological zoom--in simulations (e.g. \citealt{2019MNRAS.487.1689P}). Furthermore, it is also pivotal for quantifying the SFRD at high--redshift with an unbiased tracer, not affected by obscuration.\\
\indent The paper is organized as follows. In Sect.~\ref{sec:alp} we briefly describe the ALPINE data and the ancillary photometry.
In Sect.~\ref{sec:search} we present the search for the serendipitous lines and the fidelity and completeness assessment.
Sect.~\ref{sec:ide} is devoted to the identification of the lines. In Sect.~\ref{sec:lf} we show the \cii\ luminosity function and compare it with other observational studies and models predictions. Sect.~\ref{sec:sfrd} deals with the cosmic star formation rate density. We finally summarize the main results in Sect.~\ref{sec:con}.\\
\indent We adopt a $\Lambda$CDM cosmology using $\Omega_{\Lambda} = 0.7$, $\Omega_{\rm M} = 0.3$ and $H_0 = 70$ \kms Mpc$^{-1}$. We assumed a \citet{2003PASP..115..763C} initial mass function (IMF).

%
%
%
%
\section{ALPINE in a nutshell}
\label{sec:alp}
In this section we briefly describe the ALPINE project and the ALMA and ancillary data used in this work. Rather than being used to study the main UV--selected targets, the ALPINE datacubes were employed to look for serendipitous sources, as will be exhaustively described in Sect.~\ref{sec:search}. We call $``$serendipitous$"$ every line that is detected at a distance larger than $1"$ from the targeted UV--galaxies (see also \citealt{2020arXiv200200962B}).

\subsection{Data description}
The primary goal of ALPINE is to study the \cii\ emission in a statistical sample of galaxies \citep{2019arXiv191009517L}.
The targets are 118 UV--selected star--forming galaxies, placed on the SFR--M$_{\bigstar}$ $``$main sequence$"$ (e. g. \citealt{2011ApJ...739L..40R, 2014ApJS..214...15S, 2019arXiv191201621F}). Their redshifts are robustly constrained by UV--optical spectroscopy. 
The galaxies are located in well--studied sky regions, i.e. the Cosmic Evolution Survey field (COSMOS; \citealt{2007ApJS..172....1S}) and the Extended Chandra Deep Field-South (ECDFS; \citealt{2004ApJ...600L..93G, 2010ApJS..189..270C}).
For 75 out of 118 galaxies (64 \% of the sample) the \cii\ emission was successfully detected while only 23 sources show significant continuum emission (20 \% of the sample). For a comprehensive description of the targets catalogs see \citealt{2020arXiv200200962B}.

The observations were carried out using ALMA band 7 during Cycles 5 and 6. Two frequency settings were adopted to observe two redshift windows at $4.40 < z < 4.58$ and $5.13 < z < 5.85$.
The achieved noise is, on average, 0.14 Jy/beam \kms\ over a line width of 235 \kms and 39 $\mu$Jy/beam over the continuum.
The data reduction and processing was handled with the software \textit{CASA} (see \citealt{2020arXiv200200962B} for a full description). The visibilities were imaged using a natural weighting of the uv--plane, as the best compromise between spatial resolution and sensitivity.
We used a pixel size of $0.15"$ and an image size of $256 \times 256$ pixels in order to properly sample the primary beam ($\sim 21"$ at 300 GHz).
The final 118 datacubes have a channel width varying from 26 \kms (highest frequency setting) to 33 \kms (lowest frequency setting).
The average spatial resolution is $0.85" \times 1.13"$.
The total area covered by each pointing is 0.41 arcmin$^2$.
However, in order to guarantee an adequate sensitivity, we limited the search of the serendipitous lines to a smaller area (see Sect.~\ref{sub:fidelity} for the details). We also excluded a circle of 1$"$ radius around the phase center to avoid the emission due to the central UV--targets. This entails a final effective sky area of 27.42 arcmin$^2$ (0.23 arcmin$^2$ per pointing) where the serendipitous sources can be detected\footnote{We note that this survey area is higher than the area reported by \citealt{2020arXiv200200962B} since we included in our estimate a region where the primary beam attenuation reaches the 90\% while in \citealt{2020arXiv200200962B} the 80\% region has been considered.}.
\subsection{Ancillary photometry}
Since ALPINE observed extensively studied fields, all the sources located in the 118 pointings benefit from a wealth of multi--wavelength ancillary data (see \citealt{2019arXiv191201621F} for a comprehensive description).
The UV to near--infrared photometry is widely covered by the COSMOS15 and 3D--\textit{HST} catalogs \citep{2016ApJS..224...24L, 2012ApJS..200...13B} and \textit{HST} imaging \citep{2007ApJS..172..196K, 2011ApJS..197...36K}.
These catalogs also contain estimates of the photometric redshifts, which were used to guide the line identification (see Sect.~\ref{sec:ide}). In addition, for the two fields there are also \textit{Spitzer}--IRAC images at 3.6, 4.5, 5.8 and 8 $\mu$m \citep{2012sptz.prop90042C, 2013ApJ...769...80A, 2013ApJS..207...24G, 2007ApJS..172...86S, 2016ApJS..224...24L}, MIPS \citep{2003mglh.conf..324D, 2009ApJ...703..222L} and \textit{Herschel} data \citep{2011A&A...532A..90L, 2011A&A...533A.119E}. 
Also \textit{Chandra} data are available for the sources located in the COSMOS field \citep{2016ApJ...817...34M}.
Finally, at the longest wavelengths, deep JVLA observations at 3 GHz provide estimate of the radio continuum \citep{2017A&A...602A...1S}. The study of the spectral energy distribution (SEDs) of the serendipitous sources based on their photometric properties will be presented in another paper (Loiacono et al., in prep.).
\section{Search for the serendipitous emission lines}
\label{sec:search}
\subsection{Code description}
\label{sub:code}
We performed the search for the serendipitous lines using \emph{findclumps} (see \citealt{2016ApJ...833...69D, 2016ApJ...833...67W} for an exhaustive description), a code designed to look for sources without any prior knowledge of their frequency and spatial position, and which has been already exploited in the ASPECS survey \citep{2016ApJ...833...67W, 2019ApJ...882..138D}. In short, the algorithm performs a floating average of the channels over a range of kernels (number of channels) and searches for peaks exceeding a given signal--to--noise ratio (S/N). The latter is defined as the peak flux density as measured in the averaged map divided by the RMS computed within the entire map.\\
\indent We executed the search on the 118 ALPINE datacubes adopting a S/N threshold of 3.
For each pointing the search was repeated on datacubes of different channel width, from $\sim 90$ \kms\ to 550 \kms, since these values are compatible with the typical widths of mm--lines at high--$z$ \citep{2015Natur.522..455C, 2019ApJ...882..136A}.
The probability of a detection is indeed maximized when the channel width is of the order of the full--width at half--maximum (FWHM) of the line, while it is lowered when the channel width is larger/narrower.\\ 
\indent After the search, we removed the double detections from the output list, i.e. all the peaks at a distance lower than the beam size and in contiguous channels for each detection.
We also repeated the search after subtracting the continuum for those lines for which also its emission was detected, in order to obtain S/N referring on the line emission only (for the continuum source detection method see \citealt{2020arXiv200200962B}). Moreover, for those lines detected in datacubes with different channel widths, we considered the detection with the highest S/N as the final entry for our catalog. We obtained in this way the final list of the line candidates, where a mixture of real lines and spurious detections (i.e. noise peaks exceeding the S/N threshold) is expected.
\begin{figure}
   \centering
   \includegraphics[scale=0.35]{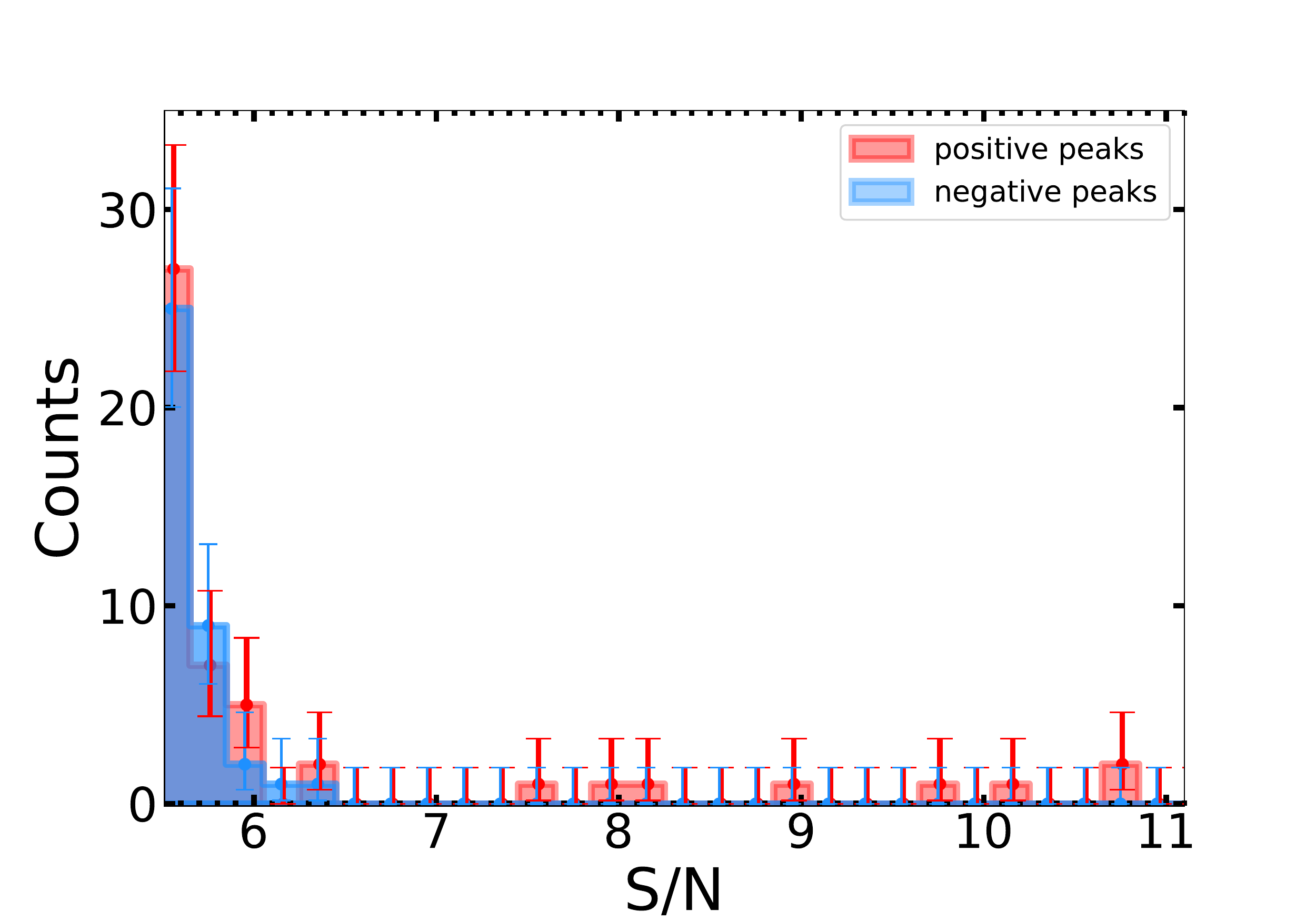}
      \caption{Number of positive (red) and negative (i.e. noise; blue) peaks detected in the 118 ALPINE pointings as a function of the S/N. The errorbars are the Poissonian uncertanties. We can see that for S/N > 5.8 the number of positive peaks becomes higher than the number of the negative ones as the number of genuine detections wrt spurious sources increases.
              }
        \label{fig:hist_peaks}
   \end{figure}
\subsection{Fidelity}
\label{sub:fidelity}
\begin{figure}
   \centering
   \includegraphics[scale=0.35]{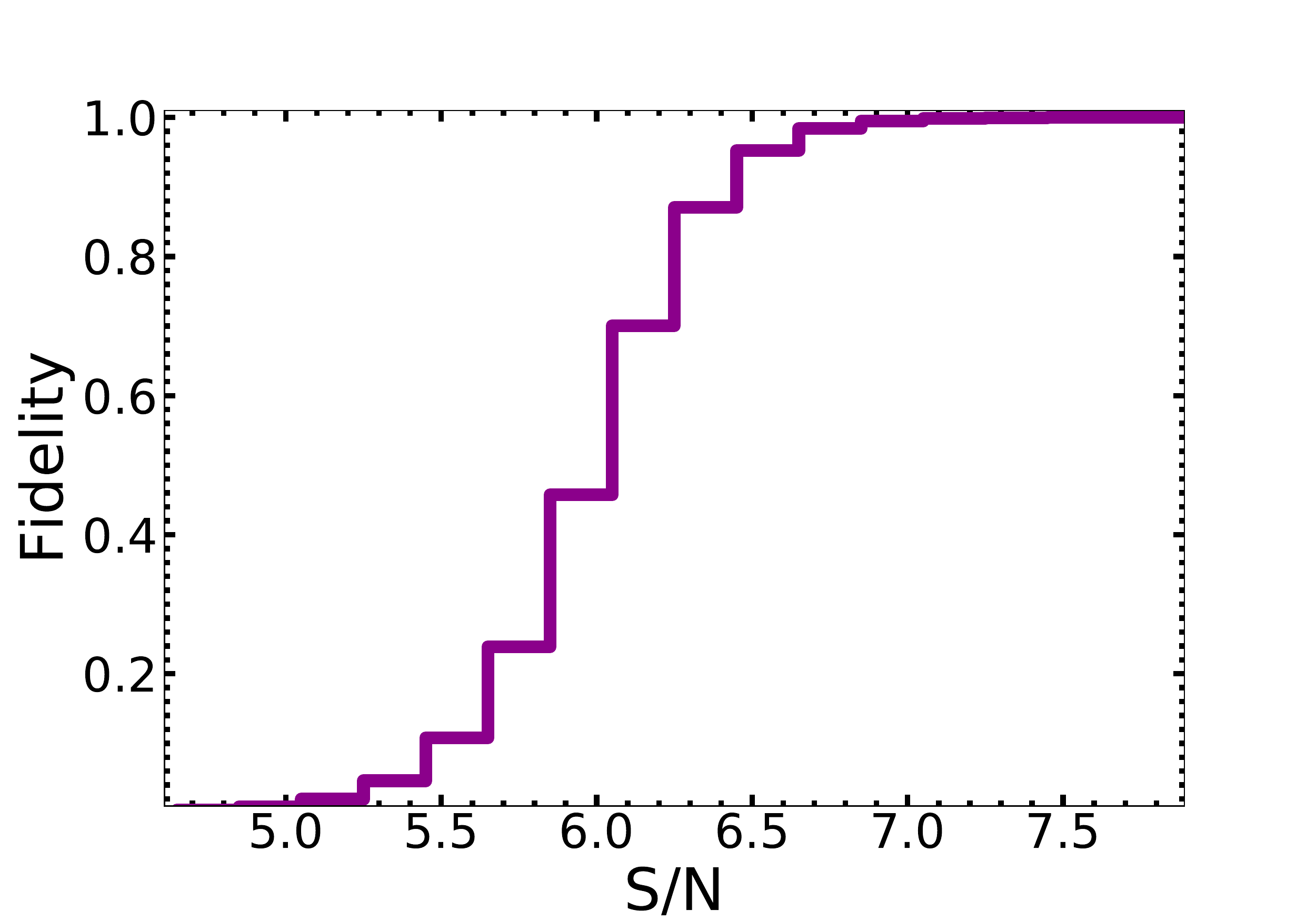}
      \caption{Fidelity curve for the serendipitous lines detected in ALPINE. The fidelity was computed by comparing the number of positive (genuine lines and noise peaks) and negative (only noise) peaks detected in the 118 ALPINE pointings. We see that the fidelity is a very steep function of the S/N. We adopted a fidelity threshold of $85$\% (corresponding to a S/N cutoff of 6.3) for the final catalog of the serendipitous lines.
              }
        \label{fig:fidelity}
   \end{figure}

In order to disentangle the genuine lines from the noise peaks in the output list, we compared the number of the positive peaks detected in the datacubes (i.e. real lines and noise peaks) with the number of negative peaks above the threshold, as a function of the S/N. Unlike the positive ones, the negative peaks provide indeed the distribution of the pure noise of our data. 
This comparison provides the fidelity, i.e. the probability that one detection is a genuine line.
Following the approach of \citet{2016ApJ...833...69D} we defined the fidelity $f$ as
\begin{equation}
{f} (S/N) = 1 - \frac{N_{\rm neg} \rm(S/N)}{N_{\rm pos} \rm(S/N)}
\end{equation}
where $N_{\rm neg}$ and  $N_{\rm pos}$ are the number of negative and positive peaks respectively.
Defined in this way, the fidelity looks like a function of the S/N only of a detection.
We note that in principle there are other factors that could influence it. For instance, the fidelity could be also a function of the line width as, for two detections of equal S/N, a larger line has a higher fidelity than a narrower one (see \citealt{2019ApJ...882..139G}). Moreover, the fidelity can also depend on the line location in the field of view (FOV), since the sensitivity within the primary beam is not uniform.  
However, because of the low statistics of the positive/negative peaks above S/N = 5.6 (below ten counts per bin even considering the 118 pointings; see Figure~\ref{fig:hist_peaks}), it was not possible to split the peaks in sub-samples based on their distance from the pointing center and their width. This S/N range is indeed crucial to assess the fidelity, as the number of genuine detections starts to be significant compared to the noise peaks at S/N$\sim 5.8$ (Figure~\ref{fig:hist_peaks}).
We thus consider only one fidelity curve, which is valid for the entire sample (Figure~\ref{fig:fidelity}). We note that the curve was computed after having excluded the peaks located in the regions with a primary beam attenuation larger than $90 \%$ as we do not expect sources at those radii. We excluded also the region within 1$"$ from the phase center to remove the positive peaks due to the central targets. The inclusion of the central targets would bias indeed the fidelity to higher values.
We note that the fidelity is very steep, jumping from 0.2 to 0.8 in a narrow range of S/N.\\
\indent We used the fidelity to define the final catalog of the serendipitous lines. We included in it the lines with a fidelity higher than 85\% (corresponding to a S/N = 6.30 cutoff). This sample includes 12 line detections. We added two more lines with lower fidelity ($\sim 50$\%, corresponding to S/N$\sim 5.98$) based on the fact that they present an optical-NIR counterpart (see Sect.~\ref{sec:ide}).
This provides a final catalog of 14 serendipitous line detections over the entire ALPINE pointings.\\
\indent We note that the adopted fidelity cut certainly excludes some genuine detections with low S/N from our catalog. Indeed, if we push down the fidelity to 20\% (S/N = 5.69), we find 10 more sources.
According to their fidelity we expect that the fraction of true sources is low ($\sim 30$\%).
However, their exclusion could have an impact on the derivation of the luminosity function (see Sect.~\ref{sec:lf}). We address this point in Sect.~\ref{sec:lf} and in Appendix~\ref{app:lf}. 
 \begin{figure*}
\centering
\includegraphics[scale=0.18]{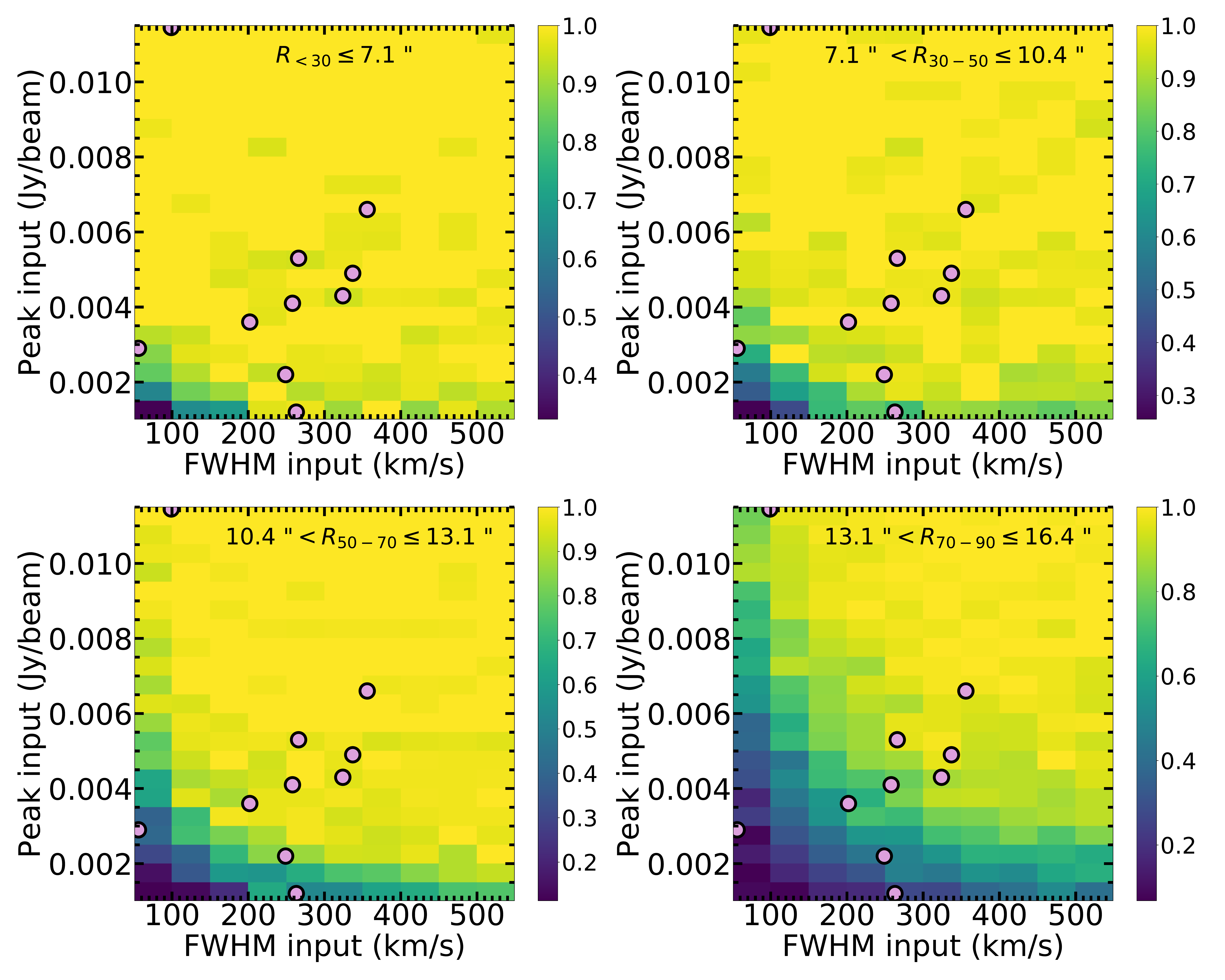}
\caption{Completeness (color scale) as a function of the flux peak and the FWHM of a line. The four diagrams correspond to the $R_{<30}$, $R_{30-50}$, $R_{50-70}$, and $R_{70-90}$ regions respectively. As it is evident from their comparison, the completeness is a strong function of the line location in the FOV because of the degrading sensitivity from the phase center to larger radii. The lines used to build the \cii\ luminosity function (see Sect.~\ref{sec:lf}) are also shown (filled circles), except for the two brightes ones (i.e., S848185 and S842313) that are located outside the plotted ranges and have completeness equal to one everywhere in the FOV. 
We show the \cii\ serendipitous detections in all the panels since we computed their completeness in each ring when building the luminosity function, independently from the line location in the FOV (see Eq.~\ref{eq:lf}).}  
\label{fig:completeness}
\end{figure*}      

\subsection{Completeness}
\label{sub:compl}
\begin{figure*}
\centering
\includegraphics[scale=0.70]{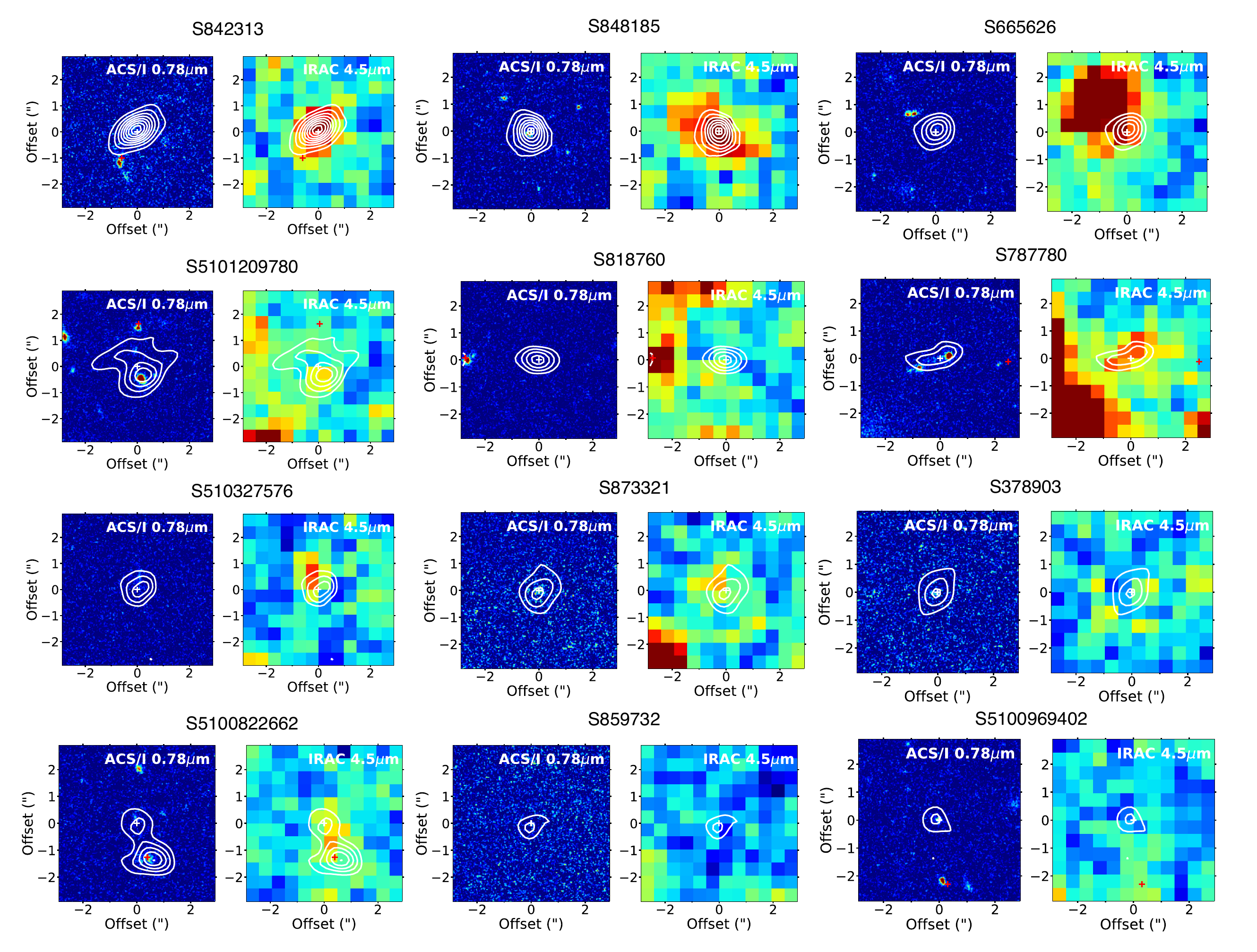}
\caption{Images cutouts of the 12 serendipitous lines used to build the \cii\ luminosity function. The $HST$-ACS 0.78 $\mu$m \citep{2007ApJS..172..196K, 2011ApJS..197...36K} and $Spitzer$--IRAC 4.5 $\mu$m \citep{2012sptz.prop90042C} are reported. The white contour shows the \cii\ emission (lowest level at 3$\sigma$). We indicate with a white cross the location of the serendipitous detection while the red cross shows the position of the central target. We can see that for 6 out of 12 lines the distance between the central target and the serendipitous line is $< 3"$ hence we are possibly witnessing interacting systems. For S5100822662 the \cii\ emission is blended with that of the central target.}  
\label{fig:cutouts}
\end{figure*}   
Given the purpose of the present work, we need to  estimate also the completeness of the sample, i.e. the fraction of recovered lines with respect to the underlying population.
We assessed the completeness by simulating $\sim 50000$ Gaussian--like lines with various peak flux $F$ and FWHM and by injecting them in datacubes containing pure noise representative of the survey (0.14 Jy/beam \kms\ over a line width of 235 \kms ). We injected the lines in random locations in the FOV and along the spectral axis, splitting them in groups of 15 lines per datacube in order to not artificially increase the source confusion. We simulated point sources ($1.16" \times 0.78"$) since the sources in our catalog are point--like or marginally resolved. However, we note that recent studies reported the existence of extended \cii\ structures (e.g., \citealt{2019ApJ...887..107F, 2020arXiv200300013F, 2020arXiv200413737G, 2020A&A...633A..90G}), which may cause incompleteness for some faint objects (see Fig. 5 in \citealt{2017ApJ...850...83F}). The simulated FWHM range is between 50 and 550 \kms , while the peak flux varies between 1.0 mJy/beam and 12 mJy/beam in order to widely sample the parameter space of the detected lines (see Figure~\ref{fig:completeness}). In particular, for each line the primary beam attenuation is taken into account, i.e. its peak flux is lowered based on the primary beam response depending on its spatial position. We hence derived the completeness $C$ in the $j-$th cell of the $(\rm FWHM, F)$ grid as
\begin{equation}
C^j({\rm FWHM}, F) = \frac{N^j_{\rm rec}({\rm FWHM}, F)}{N^j_{\rm inj}({\rm FWHM}, F)}
\end{equation}
where $N^j_{\rm inj}$ and $N^j_{\rm rec}$ are the number of injected lines and recovered lines by \emph{findclumps} in the cell.
We considered cells of 50 \kms\ and 0.5 mJy/beam width. This cell size allows us to accurately evaluate the completeness, with an average number of 60 lines in each cell. 
We note that completeness is a strong function of the line location in the FOV since the sensitivity decreases significantly as the distance from the phase center increases. We thus evaluated it locally, splitting the lines in four regions based on the primary beam response.
In particular, we defined four rings of radii $R_{<30}$, $R_{30-50}$, $R_{50-70}$, $R_{70-90}$, in which the primary beam attenuation goes from zero to the 30\% (distance from the phase center $R_{<30} \leq 7.1 "$), from 30\% to 50\% ($ 7.1 " < R_{30-50} \leq 10.4 "$), from 50\% to 70\% ($10.4 " < R_{50-70} \leq 13.1 "$) and from 70\% to 90\% ($ 13.1 " < R_{70-90} \leq 16.4 "$) and computed the completeness for each of these regions.
We avoided the separation in narrower rings since it would have implied a poor statistics of fake sources to adequately sample the completeness.\\
\indent The diagrams showing the completeness in the four rings are presented in Figure~\ref{fig:completeness}. It seems clear from the plots that for equal FWHM and peak flux, lines that are easily detected close to the phase center though become tricky to be detected when observed in the outskirts of the FOV. 
We show also the location of the lines used to build the \cii\ luminosity function (see Sect.~\ref{sub:build} and Table~\ref{table:catalog_cii}) in the parameter space (FWHM, $F$). 
All the lines have a completeness higher than 95\% in the two most internal regions except for two cases that have completeness between 90\% and 70\%. In the remaining less sensitive rings the completeness is still higher than 65\% in all the cases except for three sources with completeness values below 50\%.
This fact guaranties that we applied small completeness corrections to our lines when evaluating the luminosity function (see Sect.~\ref{sec:lf}).
We present also the completeness curves as a function of the flux peak for fixed FWHM in Appendix~\ref{app:comp}. We can see that at fixed flux peak the completeness is obviously higher for larger lines. 
\section{Identification and sources properties}
\label{sec:ide}
\begin{table*}
\caption{Catalog of the serendipitous emitters in ALPINE (confirmed and candidates; the latter are marked with a *). The sources names are labelled according to the ID number of the UV target in the same pointings, preceded by letter "S" that stands for $``$serendipitous$"$. The reported parameters were estimated using a Gaussian fit (see Appendix~\ref{app:spec}) to the line emission. Also the de--convolved sizes and the distance from the central target (in arcsec) are reported. We show also the redshift separation $\Delta z$ between the central target and the serendipitous \cii\ line in the same pointing. The only galaxy in the $``$field$"$ sample is S510327576 (see Sect~\ref{sec:lf}). The continuum flux density was measured by \citet{2020arXiv200200962B}. We report also the two CO line detections.}          
\label{table:catalog_cii}      
\centering
\renewcommand{\arraystretch}{1.5}          
\resizebox*{1.01\textwidth}{!}{	
\begin{tabular}{c c c c c c c c c c c c c c c}     
\hline\hline       
ID  & Line & S/N & Frequency & FWHM & Line flux &  Continuum flux & Optical-NIR & Ancillary &$z_{\rm line}$ & $|\Delta z|$  & $\log{L_{\rm line}}$ & Size & Fidelity & Distance\\
&&&(GHz)&(\kms)&(Jy\kms)&(mJy)&counterpart&redshift&&&(L$_{\odot}$)&(arcsec)&&(arcsec) \\ 
\hline                    
S842313& \cii & 28.18 & 343.124 & 889 $\pm$ 35 & 8.45 $\pm$ 0.29 & 8.24$\pm$0.09 &yes & spec-$z$ & 4.5389 $\pm$ 0.0001 & 0.0148 & 9.72 & $0.89" \times 0.45"$ & 1.00&1.17\\  
S848185\tablefootmark{a}& \cii & 15.97 & 301.839 & 472 $\pm$ 20 &  11.57 $\pm$  0.65 & 5.983$\pm$0.227 & yes &spec-$z$ & 5.2965 $\pm$ 0.0002 & 0.0034 & 9.96 & $0.91" \times 0.61"$&1.00&15.17\\ 
S665626& \cii * & 10.76 & 340.752& 324 $\pm$ 19 & 1.47 $\pm$ 0.12 & 0.392$\pm$0.087 & no & ... & 4.5775$\pm$ 0.0001 &0.00020&8.96& $0.66" \pm 0.46$ &1.00&6.35\\ 
S5101209780& \cii & 10.66 & 341.275& 356  $\pm$ 19 & 2.50 $\pm$ 0.18 & ... & yes & photo-$z$ & 4.5686$\pm$ 0.0001 &0.0014&9.19& $1.65" \times 1.26"$&1.00&1.64\\ 
S818760& \cii * & 10.25 & 341.450& 202 $\pm$ 12& 0.78 $\pm$ 0.06 & 0.425$\pm$0.104 & no & ... & 4.56609$\pm$ 0.00008 &0.0048&8.69&not resolved&1.00&2.73\\ 
S787780& \cii & 9.02 & 344.866& 258 $\pm$ 14 &1.13 $\pm$ 0.08& 0.398$\pm$ 0.106 & yes & spec-$z$ & 4.51095$\pm$ 0.00009 &0.00005&8.84&not resolved&1.00&2.49\\ 
S510327576& \cii & 8.14 & 355.894 & 337 $\pm$ 23 & 1.75 $\pm$ 0.16 & ... &yes & photo-$z$ & 4.3405$\pm$ 0.0002 &0.2194&9.00&$1.1" \times 0.84"$&1.00&7.15\\ 
S873321& \cii & 8.0 & 308.730& 266 $\pm$ 39 & 1.50 $\pm$ 0.29 & ... & yes & spec-$z$ &5.1560$\pm$ 0.0003 &0.0018&9.05& $1.26" \pm 0.44"$&1.00&12.69\\ 
S378903& \cii &7.5&295.858& 249 $\pm$ 26 &0.58$\pm$0.08& ... & yes& photo-$z$ &5.4238$\pm$0.0002&0.0059&8.67&not resolved&1.00&6.50\\
S5100822662& \cii *& 6.39 & 344.256& 56 $\pm$ 7 & 0.17 $\pm$ 0.03 & ... &no &...& 4.52071$\pm$ 0.00004&0.00021&8.02&not resolved&0.89&1.32\\ 
S859732& \cii *& 6.34 & 343.096& 99 $\pm$ 15 & 1.21 $\pm$ 0.24& ... &no&...& 4.5393$\pm$ 0.0001&0.0075&8.87&not resolved&0.86&12.07\\ 
S5100969402& \cii & 5.99&340.402& 263 $\pm$ 38&0.32$\pm$ 0.06& ... &yes&photo-$z$&4.5832$\pm$ 0.0002&0.0047&8.30&not resolved&0.51&2.31\\
S5110377875& CO(7--6) & 9.85 & 354.109 & 183 $\pm$ 9 & 1.35$\pm$ 0.09 & 3.512 $\pm$ 0.163 & yes & photo-$z$ & 1.27793 $\pm$ 0.00002 & ... & 7.60 & $0.88" \pm 0.59"$ & 1.00 & 6.53 \\
S460378 & CO(5--4) & 5.97 & 295.935 & 855 $\pm$ 102 & 1.11$\pm$ 0.18 & 0.680 $\pm$ 0.117 & yes & photo-$z$ & 0.9472 $\pm$ 0.0001 & ... & 7.12 & not resolved & 0.48 & 7.99 \\  
\hline                  
\end{tabular}}
\tablefoot{
\tablefoottext{a}{We note that the difference between the line flux of S848185 reported in this work and in \citet{2014ApJ...796...84R} is due the use of different apertures. This difference has a negligible impact on the luminosity function.}
}
\end{table*}
In order to identify the detected lines we cross--matched their spatial position with the entries in the COSMOS and 3D--HST photometric catalogs (Laigle et al. 2016; Brammer et al. 2012). The astrometry offsets between these catalogs and the ALMA maps are of the order of $0.1"$ (see Faisst et al. 2020). 
In addition to this, we checked for counterparts also in the SPLASH (Capak et al. 2012), UltraVista-DR4 (McCracken et al. 2012), 24 $\mu$m--selected (LeFloch et al. 2009) and 3 GHz--selected JVLA catalogs (Smolcic et al. 2017). 
Moreover, we also visually inspected the images from UV to MIR wavelegths in order to look for faint emissions not reported in the catalogs.
We classified a galaxy as a physical counterpart of a serendipitous line if their spatial distance is less than $1 "$. The choice of this value derived from the distance distribution between the serendipitous lines and all the galaxies lying within $10 "$, which clearly presents a minimum for a distance $\sim 1 "$ for all the catalogs.\\ 
\indent Based on the photometric or spectroscopic redshift available, we identified eight lines as \cii\ and two lines as CO($J_{\rm up} = 7, 5$) transitions. The remaining four detections have an ambiguous identification because of the lack of an optical/NIR or uncertain photometric redshift from ancillary data\footnote{We note that if we consider the sources with a fidelity down to 20\% nine out of the ten new sources do not show any optical/NIR counterpart. There is only one detection associated to a galaxy with a photometric redshift that makes the line emission compatible with a CO(4--3), CO(5--4) or [C I] transition.}. All the images and spectra of the serendipitous lines are reported in Appendix~\ref{app:spec}.
We refer to a future paper for an analysis of the CO emitting galaxies (Loiacono et al., in prep.), while hereafter we will focus on the \cii\ emitters and on the ambiguous lines (i.e. 12 objects in total).

\subsection{[C II] serendipitous emitters at $ 4.3 < z < 5.4$} 
\label{sub:cii}
We identified eight lines as \cii\ based on the photometric or spectroscopic redshift of the optical/NIR infrared counterpart available from ancillary data. Namely, four out of eight detections have an UV--optical spectroscopic redshift (M. Salvato, private communication; \citealt{2008ApJ...681L..53C, 2011Natur.470..233C}). The remaining four sources have photometric redshifts compatible with \cii\ emission \citep{2016ApJS..224...24L}. 
The sources have redshift $4.3 < z < 5.4$, as expected due to the spectral coverage of ALPINE.
We note that among the serendipitous \cii\ emitters we recovered the well--studied sub--mm galaxies AzTEC--C17 (here called S842313; \citealt{2016ApJS..224...24L, 2008ApJ...689L...5S, 2017ApJ...850..180J}) and AzTEC--3 (S848185; \citealt{2011Natur.470..233C, 2010ApJ...720L.131R, 2014ApJ...796...84R}).\\
\indent 
In addition to these eight detections we found four lines whose identification based on the available photometry is ambiguous. Two of them (S818760 and S859732) do not present any counterpart in the available catalogs and also in the multi--wavelenght images (from UV to MIR).
The lack of counterparts suggests that these emissions are produced from highly dusty/high-$z$ sources or from gas--rich galaxies with low stellar masses. The most likely associations are thus \cii\ at $4 < z < 6$ or CO transitions at lower redshifts.
However, S818760 is located within $3"$ and has a velocity separation $< 300$ \kms\ from the central target in the same pointing (see Figure~\ref{fig:cutouts}). As a consequence, it is produced very likely by a companion/interacting source with the UV--target emitting also \cii\ but optically faint (see also \citealt{2020MNRAS.491L..18J}). A similar argument applies to S5100822662\footnote{We note that this particular source was included also in the luminosity function of Yan et al. 2020, as they used the total flux of the central target vuds cosmos 5100822662 (see also Figure~\ref{fig:cutouts}), which was not deblended from the serendipitous companion (see \citealt{2020arXiv200200962B}). However, since the latter respects the criterion to be a non--target source (i.e., distance from the phase higher than 1$"$) we included it in our calculations. We note that its exclusion from our sample does not alter the result significantly.}. Differently from S818760, this source is detected in the available images (see Figure~\ref{fig:cutouts}) and based on the \citet{2016ApJS..224...24L} catalog it has a photometric redshift of 0.69. However, the strict association with the ALPINE target in the same pointing (Figure~\ref{fig:cutouts}) favours a high--$z$ interpretation for this source with the ALMA emission likely due to the \cii\ line. It is also possible that the emission in the photometric images is produced by a foreground source that is not related to the ALMA detection (see also \citealt{2018ApJ...861...43P}).
Finally, S665626 does not present any counterpart in all but K--band UltraVista image \citep{2020arXiv200200961R}. This source was studied in detail by \citet{2020arXiv200200961R} and their modelling seems to favour a \cii\ interpretation than a CO line.
Follow-up observations are necessary to unambiguously confirm the nature of these four sources.\\ 
\indent In the rest of this work we assume that the four unidentified lines are due to \cii\ emission. We used both them and the confirmed \cii\ to build the luminosity function (Sect.~\ref{sec:lf}). We note that the exclusion of the unidentified lines from it does not alter significantly any of the results. The optical/NIR images of the 12 serendipitous \cii\ lines (confirmed and candidates) are shown in Figure~\ref{fig:cutouts}.\\
\indent We estimated the main properties of the \cii\ lines (i.e. frequency, FWHM, total fluxes) by performing a single--component Gaussian fit to the continuum subtracted spectrum, with the exception of source S842313 where two Gaussians were adopted to model the line profile as it shows signs of rotation \citep{2017ApJ...850..180J}.
To compute the line flux we used the peak flux if the source size is comparable with the beam or we extracted it from a 3$\sigma$ aperture in case the emission is resolved.
To distinguish between resolved and unresolved sources, we compare the number of pixels within a 3$\sigma$ aperture with the beam size in pixels. In case the number of pixels exceeds the beam size we labelled the source as resolved. Otherwise we considered the source as not resolved.
Then we evaluated the deconvolved sizes of the resolved sources using the 2D fitting tool of \textit{CASA}.
We also measured the line fluxes on the moment zero maps but we do not report them since they show consistent results. All the fitted values are reported in Table~\ref{table:catalog_cii}.\\ 
\subsection{Overdensities around the central targets}
\label{sub:over}
The detection of eight confirmed \cii\ lines in targeted \cii\ observations of $4< z < 6$ galaxies suggests that we are witnessing possible overdensities around the central UV--selected galaxies. This is highlighted from the velocity separation $\Delta v$ between the central target and the serendipitous line in the same pointing. Seven out of eight \cii\ have indeed $|\Delta v| < 750$ \kms, corresponding to a redshift separation $|\Delta z| < 0.0154$.
Such a velocity difference suggests that the two galaxies in the same pointing could be physically connected and/or associated to the same large--scale structure.
An extended protocluster at $z\sim 4.5$ (PCI J1001+0220) in the COSMOS field was discovered by \citet{2018A&A...615A..77L}. \citet{2011Natur.470..233C} found another protocluster of galaxies in COSMOS at higher redshift ($z \sim 5.3$). In fact some of the serendipitous lines in our sample (e.g. S848185) are well known members of these protoclusters.  
However there are other detections in our catalog that could constitute potential new members of these overdense regions.
This is likely valid for one confirmed \cii\ emitter and two \cii\ candidates that lie in the spatial region corresponding to PCI J1001+0220 and have a redshift in the range $4.53 < z < 4.6$ \citep{2018A&A...615A..77L} while other three \cii\ lines (two candidates and one confirmed) are possibly located in the outskirts of the same protocluster.\\ 
\indent Besides the low velocity/redshift separation, we also see that for four out of eight confirmed \cii\ the spatial separation from the central target is less then $3"$, corresponding to a physical distance $<20$ kpc at $z\sim 5$ (Figure~\ref{fig:cutouts}). The number of sources increases to six if we include also the \cii\ candidates S818760 and S5100822662. These sources are galaxies likely interacting with the central targets.  
This is also suggested from the \cii\ morphologies, which appear irregular in some cases (see for example S5101209780, Figure~\ref{fig:cutouts}; \citealt{2020arXiv200413737G}). Therefore, we could be in presence of two kind of overdensities: one on very small scale ($< 20$ kpc) due to galaxy pairs and/or mergers and another on a larger scale (up to $\sim 90$ kpc, i.e. the maximum distance allowed by the size of our pointings), related to a more extended structure. We will analyze in detail the overdense environment in a future paper (Loiacono et al., in prep.). The effect of clustering was taken into account when building the \cii\ luminosity function (see Sect.~\ref{sec:lf}).
\subsection{Relation between [C II] luminosity and SFR}
\label{sub:cont}
\begin{figure}
\centering
\includegraphics[scale=0.35]{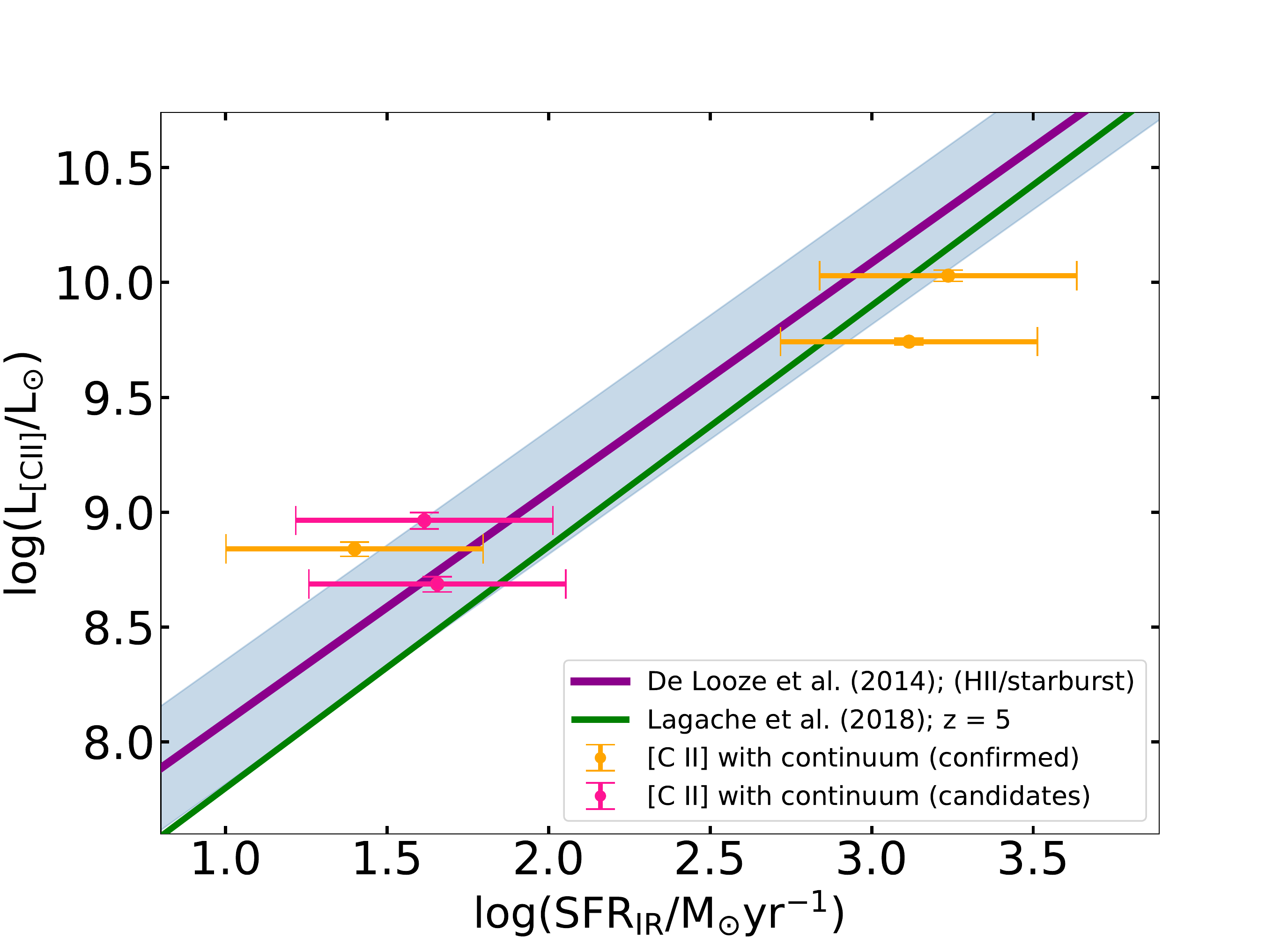}
\caption{SFR--$L_{\rm \cii}$ relation for the 5 serendipitous lines detected in continuum. The SFR was computed from the IR luminosity of the sources. We can see that the sample is quite consistent at 1$\sigma$ (colored area) with the \citet{2014A&A...568A..62D} relation (purple line), suggesting that the contribution from the UV--traced SFR is negligible. We compare our results also with the models of \citet{2018A&A...609A.130L} at $z = 5$ that suggest a slightly different slope.}
\label{fig:delooz}
\end{figure}

Within the sample of the serendipitous \cii\ lines there are five sources for which also the continuum has been detected (\citealt{2020arXiv200200962B}; Gruppioni et al. 2020).
It is well known that there is a correlation between the \cii\ luminosity and the SFR \citep{2014A&A...568A..62D}.
Since the latter is well--traced by the total IR luminosity (8-1000 $\mu$m), we used the five lines to test if this relation is valid also at $z \sim 5$.
We included also the two unconfirmed \cii\ for which the ALMA continuum has been detected (S818760 and S665626).\\
\indent The \cii\ fluxes were converted to luminosities using Eq. 1 of \citet{1992ApJ...398L..29S} and we propagated the errors from the fitted quantities in Table~\ref{table:catalog_cii}.
The total IR luminosity $L_{\rm IR}$ was estimated from a SED fitting of the galaxies. We assumed the template of a star--forming galaxy that reproduces most of the \textit{Herschel} galaxies at $z \sim 2 - 3$ \citep{2013MNRAS.432...23G}. We note that the uncertainty on the total IR luminosity can be up to a factor of 5 depending on the assumed dust temperature (see \citealt{2017ApJ...847...21F, 2020arXiv200410760F}). This uncertainty accounts for about a factor $\sim 2.5$ on the derived SFR, which we assumed as the typical error of this quantity. 
Then the $L_{\rm IR}$ was converted to SFR using the \citet{1998ApJ...498..541K} relation.\\
\indent If we compare our values with the local relation of \citet{2014A&A...568A..62D} (see their Table 3, case HII/starburst) we can see that they are broadly consistent within the 1$\sigma$ errorbars (Figure~\ref{fig:delooz}).
On the other hand, our points suggest a slightly different slope compared to the model predictions at $z = 5$ of \citet{2018A&A...609A.130L}. 
The same trend is also shown by the ALPINE targets \citep{2020arXiv200200979S, 2020arXiv200200962B}, which do not present any evidence of evolution of the SFR--$L_{\rm \cii}$ relation between $z\sim 0$ and $z\sim 5$. The only difference is that, compared to the central UV--galaxies, the serendipitous sources sample a different space in the SFR--$L_{\rm \cii}$ plane, shifted to higher SFRs and \cii\ luminosities.
Besides, we note that the SFR of the ALPINE targets takes into account both UV and IR estimates, otherwise the UV--targets would not lie on the \citet{2014A&A...568A..62D} relation (see \citealt{2020arXiv200200979S} for the details).
On the other hand we considered the IR--derived SFR only for the serendipitous galaxies.
This means that the serendipitous sources detected both in line and continuum have SFRs dominated by the IR emission, with little or negligible contribution from UV.\\
\indent Therefore both the ALPINE targets and the serendipitous sources seem to independently suggest that there is no significant variation of the relation of \citet{2014A&A...568A..62D} up to $z \sim 5$. 
\begin{figure}
\centering
\includegraphics[scale=0.6]{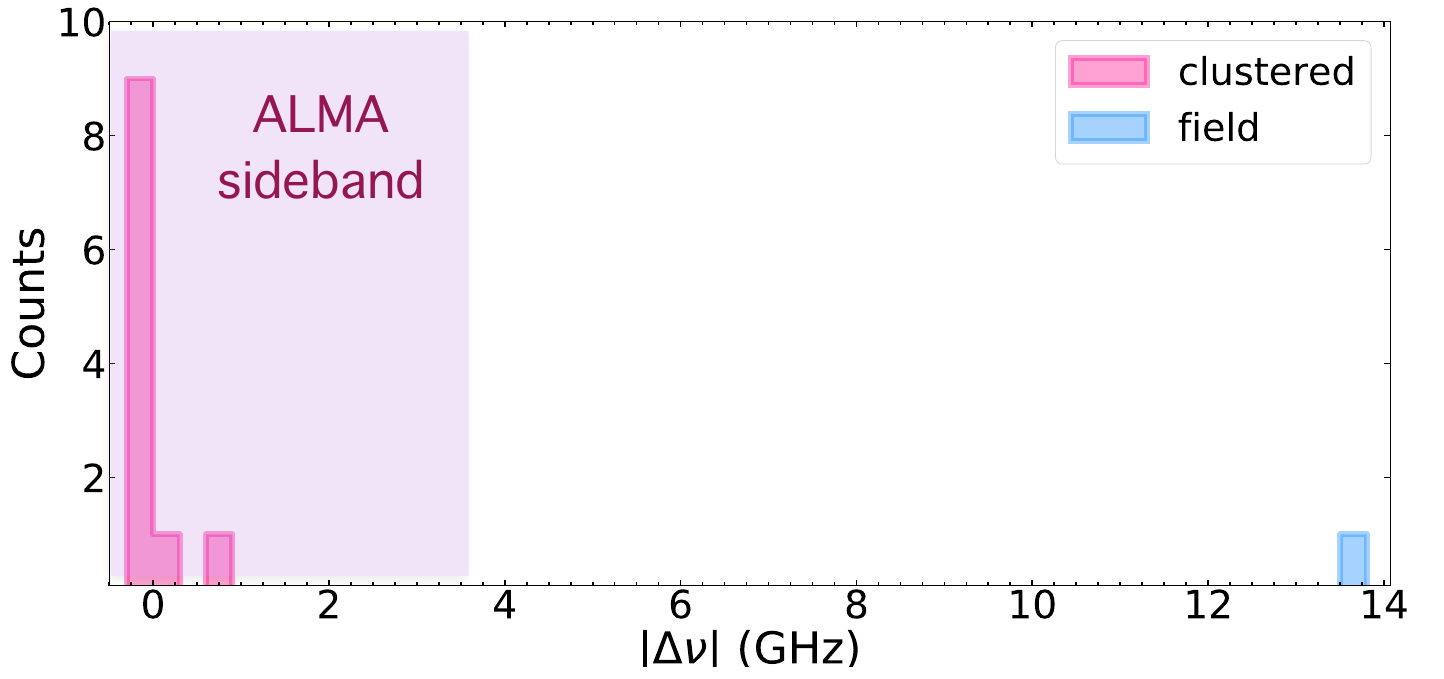}
\caption{Offset in frequency between the central UV target and the serendipitous \cii\ in the same pointing. We see that the distribution is non--uniform, with several sources lying at a frequency (and hence a redshift) close to that of the central target. 
We thus defined two sub--samples (named $``$clustered$"$ and $``$field$"$ respectively) and evaluated two distinct LFs in order to account for any bias due to overdense regions. The separation between the two sample relies on the frequency width of one ALMA sideband (3.6 GHz), corresponding to a velocity separation $\gtrless 2000$\kms .}
\label{fig:hist_deltaz}
\end{figure}
\section{The \cii\ luminosity function at $z\sim 5$}
\label{sec:lf}
\subsection{Building of the luminosity function}
\label{sub:build}
\begin{figure*}
\centering
\includegraphics[scale=0.6]{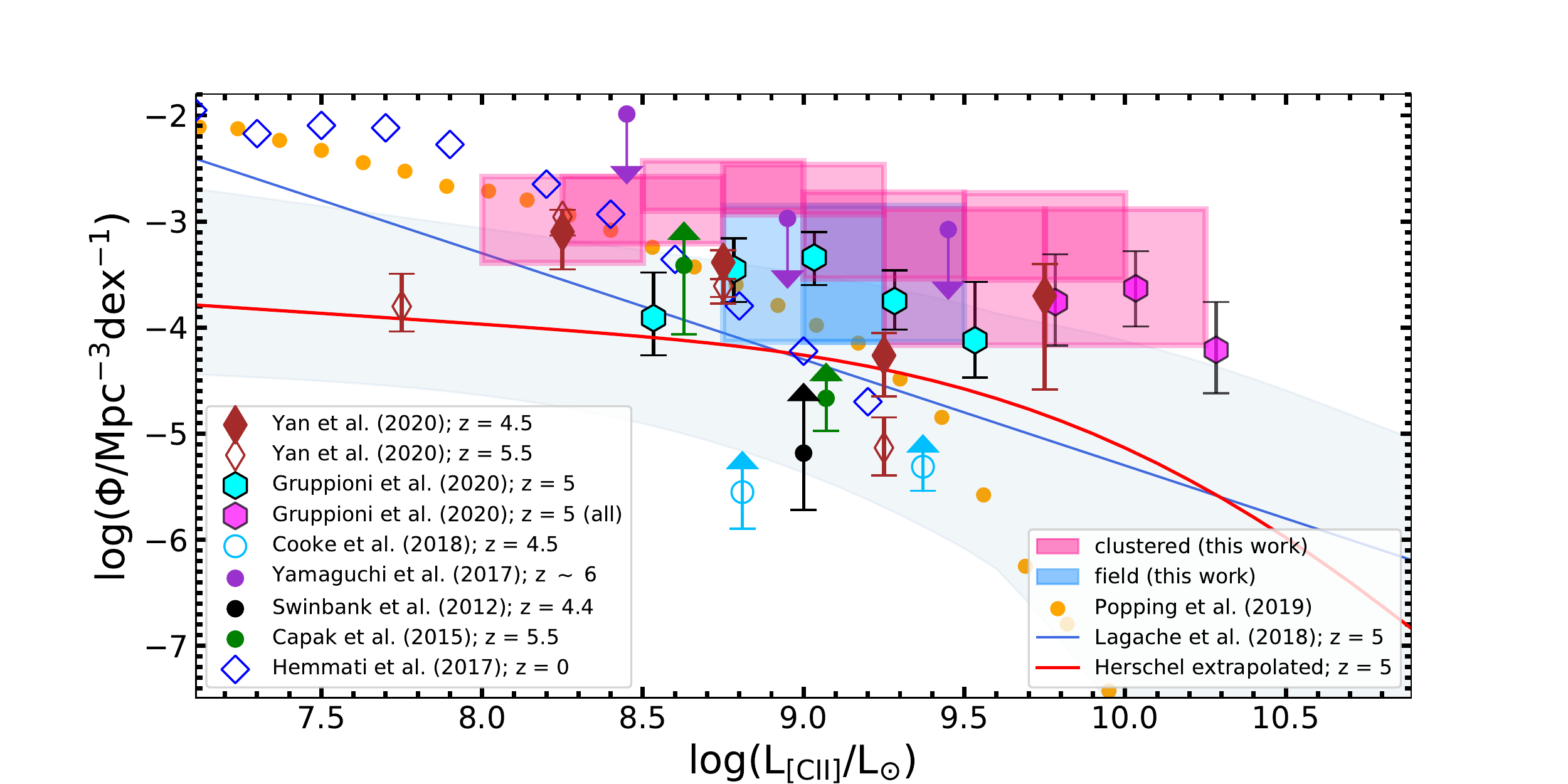}
\caption{\cii\ luminosity functions (LFs) at $z \sim 5$ from the serendipitous sources in ALPINE compared to other works in the literature. We splitted the lines in two sub--samples, called $``$clustered$"$ (pink) and $``$field$"$ (azure) respectively, and built two separate LFs. Compared to the clustered LF, the field one lies at lower luminosities. We compare our \cii\ LFs at $z \sim 5$ with other \cii\ LFs at high and low-$z$. Overall, the estimates from the clustered sample lie above the LFs of the ALPINE targets (Yan et al. 2020) likely because they include also UV--dark galaxies and because of the clustering effect. On the other hand the field LF seems to be quite consistent with the targets ones except for the highest luminosity bin. 
There is a nice agreement between the field \cii\ LF and the IR--derived \cii\ LF based on the ALPINE serendipitous sources detected in continuum (Gruppioni et al. 2020). The agreement persists at $L_{\rm{\cii }} > 10^{9.5} {\rm L}_{\odot}$ for the clustered sample if the companions of the central targets are included in the IR--derived \cii\ LF of Gruppioni et al. (2020). The clustered LF is up to > 1 dex higher than the local \cii	\ LF \citep{2017ApJ...834...36H}. Also the field LF predicts an excess of \cii\ emitters at $L_{\rm \cii} > 10^{9} {\rm L}_{\odot}$, suggesting a possible evolution of the \cii\ LF between $z\sim 5$ to $z\sim 0$. 
The field LF appears in agreement with the models predictions of \citet{2019MNRAS.482.4906P}.}
\label{fig:lf_models}
\end{figure*}

The 14 \cii\ lines (eight confirmed and four candidates) were used to build the luminosity function (LF). 
We populated the $i$-th luminosity bin dlogL$_i$ according to the relation
\begin{equation}
\Phi(L_i) \text{d}\log{L_i} =  \Sigma_j \frac{F_j}{\Sigma_k C_j^k V^k}
\label{eq:lf}
\end{equation}
where $\Phi(L_i) {\rm{d} \log{L_i}}$ is the number density of \cii\ emitters, $F_j$ and $C_j^k$ are the fidelity and completeness of the $j$-th source associated to the comoving volume $V^k$. The latter was evaluated for the regions $R_{<30}$, $R_{30-50}$, $R_{50-70}$, $R_{70-90}$ in order to take into account the completeness variation in the FOV, hence the $k$ index goes by the four rings. Only the sources with completeness and fidelity equal to unity everywhere in the FOV would have been indeed observable within the total comoving volume $V_{\rm TOT}$ covered by the 118 ALPINE pointings. This volume was evaluated as $V_{\rm TOT} = \Sigma_i^{118} A_i \Delta D_c(z_i) = 9810\ {\rm Mpc}^3$ where $A_i$ is the area with a primary beam attenuation < 90\% covered by each ALPINE pointing and $\Delta D_c(z_i)$ is the difference between the comoving distances of the \cii\ line at the beginning and at the end of the ALMA sidebands for the $i$-th pointing. 
This difference was computed after having excluded 3--4 channels at the beginning and at the end of each sideband to account for border effect (i.e. noisy channels). 
We note that we excluded the central $R < 1"$ region from each pointing in the computation of the volume.
As the luminosity bin size we considered 0.5 dex in order to have at least one source per bin.
The adopted bin spacing is 0.25 dex in luminosity. Although the bins are not independent this choice gives the advantage to better highlight the luminosity distribution of the sample. 
We point out that we did not split the \cii\ lines in different redshift bins because of the poor statistics, hence our LF refers to an average redshift $z \sim 5$. As done in Sect.~\ref{sub:cont}, we evaluated the \cii\ luminosities following \citet{1992ApJ...398L..29S} (see Table~\ref{table:catalog_cii} for the values).
The errorbars associated to each luminosity bin are computed as the Poissonian uncertainties corresponding to 1$\sigma$ since the source number in each bin is small \citep{1986ApJ...303..336G} and thus constitutes the major uncertainty.\\
\indent Before computing the LF, we splitted the \cii\ lines in two sub--samples. As we saw in Sect.~\ref{sec:ide}, seven out of eight confirmed \cii\ have a redshift separation from the central targets in the same pointings $|\Delta z| < 0.015$  (corresponding to a velocity separation $< 750$ \kms). This number increases to 11 out of 12 if we include also the four unconfirmed \cii . This means that their LF could be not representative of the field galaxy population since it is likely biased by the presence of overdensities around the UV--selected targets.
The only exception is S510327576, which has a redshift separation $|\Delta z| = 0.2195$ ($|\Delta v| \sim 1.2 \times 10^4$ \kms) and thus is not related to the central target.
This could be the only \cii\ line not associated to clustered structures, i.e. the only genuine field source in our sample.\\
\indent In order to study the effect of clustering on the LF, we thus considered two separate sub--samples, each of them containing the lines with a frequency offset from the central target lower/higher than one ALMA sideband ($\Delta \nu \sim 3.6$ GHz; see Figure~\ref{fig:hist_deltaz}). 
This separation corresponds to a redshift difference $\gtrless 0.04$ and to a velocity separation $\gtrless 2000$ \kms (see also \citealt{2010ApJ...719.1672H} who used a similar velocity separation in a study on quasars pairs).
In this way we defined the $``$clustered$"$ and $``$field$"$ sub--samples, containing 11 and one sources respectively. Also the survey volume was splitted consistently, obatining a total comoving volume of $5026\ \rm Mpc^3$ for the clustered sub--sample and $4784\ \rm Mpc^3$ for the field one.
Thus we built a separate \cii\ LF for each sub--sample (Figure~\ref{fig:lf_models} and Table~\ref{table:lf_id}).
The median luminosity of the clustered sample ($\log{(L/\rm L_{\odot})}= 8.96 \pm 0.14$) is very similar to the luminosity of the field one ($\log{(L/\rm L_{\odot})}\sim 9.0$). However, we remind that the field LF is based on one object only and therefore it could present also galaxies at higher luminosity that we do not detect for the limited survey volume.
Despite the similar median luminosity, the clustered LF shows objects with luminosity of about one order of magnitude higher than the field.
If this trend were confirmed by a larger sample of galaxies it would highlight a dependence between clustering and the \cii\ luminosity, as already shown based on other tracers (e.g. \citealt{2001MNRAS.325..589H}).\\
\indent In Appendix~\ref{app:lf} we report also the LF computed including sources with fidelity as low as 20\% (see Figure~\ref{fig:lf_dook}). 
In Sect.~\ref{sub:fidelity} we cut indeed our catalog of serendipitous detections at a fidelity of 85\%, with only one \cii\ line (S5100969402) having a fidelity $\sim 50$\%, in order to study a very robust sample. However this sample is obviously incomplete at low luminosity. 
We thus calculated the clustered and field LFs for two new sub--samples, in which we included also low fidelity (i.e. low luminosity) sources. The fidelity cut of 20\% adds nine lines to our catalog of \cii\ candidates (we excluded the only source possibly associated to CO emission; see Sect.~\ref{sec:ide}). We note that none of these lines presents an optical/NIR counterpart hence their redshift is unconstrained from ancillary data.
Therefore, for these sources we can only assume that their emission is due to \cii . 
We see that the shape of both clustered/field LFs remains quite unchanged, with the field LF sampled by more sources now. Also in this case the field sources lie at lower luminosities compared to the clustered sample. However, this is not surprising since, as the fidelity lowers, also the line flux decreases and hence we expect the population of the low luminosity bins only for both field and clustered sources.\\
\indent Finally, in Sect.~\ref{sub:over} we saw that the clustered sources are possibly part of two different types of overdensity, one associated to interactions/mergers (scale $< 20$ kpc) with the central UV--selected galaxy and the other associated to a more extended structure (up to $\sim 90$ kpc). 
In order to overcome the bias introduced by the interacting systems, we excluded from the LF the 6 sources with a spatial distance $< 3"$ from the central target (Figure~\ref{fig:cutouts}). 
We report the derived LF in Appendix~\ref{app:lf}.
The new points are consistent within the errors with the LF computed using all the clusterd sources.
Overall the faint end of the LF results lower compared to the case in which all the 11 clustered \cii\ are considered. However, this has a negligible effect on the derivation of quantities like the fitted parameteres of the Schechter function and the star formation rate density (see Sect.~\ref{sub:schech} and Sect.~\ref{sec:sfrd}). 
\begin{table}
\caption{Luminosity functions for the clustered and field sample considering the eight confirmed and four candidates \cii . We reported also the number of sources in each luminosity bin. We indicated with the bold font the values corresponding to independent luminosity bins.}             
\label{table:lf_id}      
\centering          
\resizebox*{0.47\textwidth}{!}{	
\begin{tabular}{c c c c c}     
\hline
\addlinespace[0.2cm]
$\log{(L/\rm L_{\odot})}$&$\log{(\Phi_{\rm clust}/\rm Mpc^{-3} \rm dex^{-1})}$& $N_{\rm clust}$ &$\log{(\Phi_{\rm field}/ \rm Mpc^{-3} \rm dex^{-1})}$ & $N_{\rm field}$ \\
\addlinespace[0.2cm]
\hline
\addlinespace[0.2cm]
 8.25& \textbf{$\textbf{-2.94}^{\textbf{+0.36}}_{\textbf{-0.45}}$} & \textbf{2} & ... & ... \\
\addlinespace[0.2cm]
8.50 & $-2.87^{+0.29}_{-0.34}$ & 3 & ... & ... \\
\addlinespace[0.2cm]
8.75& \textbf{$\textbf{-2.65}^{\textbf{+0.22}}_{\textbf{-0.24}}$} & \textbf{5} & ... & ... \\
\addlinespace[0.2cm]
9.00& $-2.69^{+0.22}_{-0.24}$ & 5 & $-3.37^{+0.52}_{-0.77}$ & 1 \\
\addlinespace[0.2cm]
9.25& \textbf{$\textbf{-3.09}^{\textbf{+0.36}}_{\textbf{-0.45}}$} & \textbf{2} & \textbf{$\textbf{-3.37}^{\textbf{+0.52}}_{\textbf{-0.77}}$} & \textbf{1} \\
\addlinespace[0.2cm]
9.50& $-3.40^{+0.52}_{-0.77}$ & 1 &  ... & ... \\
\addlinespace[0.2cm]
9.75& \textbf{$\textbf{-3.10}^{\textbf{+0.36}}_{\textbf{-0.45}}$} & \textbf{2} & ... & ... \\
\addlinespace[0.2cm]
10.00& $-3.40^{+0.52}_{-0.77}$ & 1 & ... & ... \\
\addlinespace[0.2cm]
\hline       
\end{tabular}}
\end{table}

\subsection{Comparison with observations and models}
\subsubsection{Luminosity functions from ALPINE}
In this section we discuss our LFs in relation to those from other works (Figure~\ref{fig:lf_models}).\\ 
\indent We start by comparing our results with the other $z \sim 5$ LFs based on the ALPINE data.
First of all, we consider the \cii\ LFs presented in the companion paper of Yan et al. (2020). These LFs were built using the 75 \cii\ central UV--targets in the two redshift ranges $4.40 < z < 4.58$ and $5.13 < z < 5.85$. 
Globally, we see that the clustered LF predicts more \cii\ emitters than the Yan et al. (2020) sample. This was expected due to clustering effects and also because the LF of the central targets is based on UV--selected galaxies, hence it is likely missing the most obscured galaxies.
On the other hand the field LF is quite consistent with the targets LFs, showing a slight excess in the highest luminosity bin.\\
\indent Then we compare our sample with the LF based on the sources serendipitously detected in the rest--frame FIR continuum (Gruppioni et al. 2020, \citealt{2020arXiv200200962B}). The 118 ALPINE pointings revealed indeed a wealth of serendipitous continuum emitters in a wide range of redshifts. These sources were used to build a rest--frame 250 $\mu$m LF and a total IR LF from $z = 0.5$ to $z = 6$ (see Gruppioni et al. 2020 for the details). For our comparison we considered the IR LF in the highest redshift interval $4.5 < z < 6$, where the companions of the central targets have been removed (green water hexagons; see Table 2 of Gruppioni et al. 2020).
The IR luminosities (8-1000 $\mu$m) were first converted to SFRs according to the \citet{1998ApJ...498..541K} relation. We note that the computed SFRs do not include the UV contribution, therefore they can be considered as lower limits. However we do not expect the UV contribution to be significant since the sources are selected to be dusty (i.e., FIR/sub--mm emitters).
The SFRs were then used to derive the \cii\ luminosities following the \citet{2014A&A...568A..62D} relation (case HII/starburst), scaled for a \citet{2003PASP..115..763C} IMF. 
Globally, the clustered LF presents a higher number density (up to about 1 dex) and higher luminosity objects than the IR--derived \cii\ LF of Gruppioni et al. (2020). The difference in the lower luminosity bins is however enhanced by the fact that these bins are strongly incomplete in the continuum survey (see \citealt{2020arXiv200200962B}).
On the other hand, there is perfect agreement between the field LF  and the LF derived from Gruppioni et al. (2020). 
However, if we show the IR--derived \cii\ LF that includes also the companions of the central targets for $L_{\rm{\cii }} > 10^{9.5} {\rm L}_{\odot}$ (magenta hexagons; see Gruppioni et al. 2020) we find that, in this luminosity range, the clustered \cii\ LF and the IR--derived \cii\ LF are nicely consistent within the errorbars. This is due to the fact that these luminosity bins include the same sources, clustered around the central targets, detected both in line and in continuum. 
\subsubsection{Observed luminosity function at high and low-$z$}
Now we can move on to comparing our results to other works in the literature, at high and low-$z$.
We see that our LFs are consistent with previous estimates at $z = 4.4$  and $z \sim 5$ from \citet{2012MNRAS.427.1066S} and \citet{2015Natur.522..455C}.
\citet{2012MNRAS.427.1066S} started from an original 870 $\mu$m selection of galaxies with LABOCA \citep{2009ApJ...707.1201W} and considered the only two galaxies for which the \cii\ line was detected in a subsequent ALMA follow--up. However, the low continuum detection rate of the ALPINE targets (20\%; \citealt{2020arXiv200200962B}) compared to the line detection rate (64\%) suggests that a considerable fraction of \cii\ emitting galaxies can be missed when starting from continuum pre--selected samples, hence the LF of \citet{2012MNRAS.427.1066S} likely provides a lower limit to the number density of the \cii\ emitters. In case of the estimate from \citet{2015Natur.522..455C} we use the value reported in \citet{2017ApJ...834...36H}. Also in this case the data likely provide a lower limit to the true distribution, since the targets of \citet{2015Natur.522..455C} are Lyman break galaxies, i.e. UV--selected objects, and hence \cii --bright but optically--faint objects are not taken into account in this LF. Moreover, in this estimate the \cii\ serendipitous emitters in the ten pointings of \citet{2015Natur.522..455C} are not considered (e.g. AzTEC--3, \citealt{2010ApJ...720L.131R}; CRLE, \citealt{2010ApJ...720L.131R}).\\
\indent Our values are also above the LF of \citet{2018ApJ...861..100C}. This study considers \cii\ emitting galaxies pre--selected based on their SCUBA2 850 $\mu$m flux density \citep{2017MNRAS.465.1789G}, hence also this estimate provides a lower limit.\\
\indent We also compared our estimates with measurements at higher redshift \citep{2017ApJ...845..108Y}. The points of \citet{2017ApJ...845..108Y} represent upper limits to the \cii\ LF at $z \sim 6$.
We can see that the field LF is well consistent with the upper limits.
On the other hand the clustered LF seems to predict more \cii\ emitters than \citet{2017ApJ...845..108Y} at $L_{\rm{\cii }} = 10^{8.75} {\rm L}_{\odot}$ probably because it is biased to an overdense environment.\\
\indent It is interesting to compare our work also with an extrapolation of the \textit{Herschel} LF at $z\sim 5$ (\citealt{2013MNRAS.432...23G}; Gruppioni et al., in prep.). The extrapolation was performed using the SCUBA2 number counts \citep{2017MNRAS.465.1789G} to constrain the evolution at high--redshift \citep{2019MNRAS.483.1993G}. The IR luminosities were thus converted to SFRs using the \citet{1998ApJ...498..541K} relation and the SFRs were transformed in \cii\ luminosities following \citet{2014A&A...568A..62D}. 
We note that the same approach has been already used for deriving the CO LF in \citet{2016MNRAS.456L..40V}, which successfully reproduces the observed CO LF of ASPECS \citep{2019ApJ...882..138D}. 
Interestingly, we see that the global shapes of the clustered LF and the \textit{Herschel}--derived one are in good agreement, with both LFs predicting \cii\ emitters with very high luminosities ($L_{\rm{\cii }} > 10^{9} {\rm L}_{\odot}$), with at least some of the discrepancy coming from the fact that the \textit{Herschel} extrapolation was not intended to account for the clustering inherent in the ALPINE serendipitous sample.\\
\indent Finally, we discuss how the $z\sim 5$ \cii\ LF compares with the $z \sim 0$ values \citep{2017ApJ...834...36H} to underline potential evolutionary effects. 
We can see that the clustered LF shows a strong evolution both in number density (up to $> 1$ dex) and in luminosity between $z \sim 0$ and $z \sim 5$.
The field LF suggests also a possible excess of objects at $L_{\rm \cii} > 10^{9} {\rm L}_{\odot}$ compared to the local value. The two LFs are however consistent within 2$\sigma$.
A higher statistics for the field sample is necessary to draw robust conclusions about any evolutionary trend that is independent from clustering.  
\subsubsection{Theoretical predictions}
We also compare our results with model predictions for the early Universe. First of all, we considered the models for the \cii\ LF by \citet{2019MNRAS.482.4906P}.
These are semi--analytical models that include radiative transfer modelling. 
We can see that the clustered \cii\ LF predicts a higher number of objects than the models expectations at $z\sim 5$, with a disagreement that rises with increasing luminosity. 
A similar disagreement with models predictions is seen also for the CO LFs at high--$z$ \citep{2019ApJ...872....7R} and for the IR LF at $z\sim 2$ \citep{2015MNRAS.451.3419G}.
On the other hand, the field LF appears quite consistent with the models.
Further statistics would be useful to constrain the bright end of the field LF and disentangle if it remains flat at $L_{\rm \cii} > 10^{9} {\rm L}_{\odot}$ (as for the clustered sample) or if it declines as shown by models.\\ 
\indent Then we examine the predictions at $z\sim 5$ by \citet{2018A&A...609A.130L}. This is also a semi--analytic model combined with a photoionization code. 
We note that at luminosities between $10^9 \rm L_{\odot}$ and $10^{10.5} \rm L_{\odot}$ the \citet{2018A&A...609A.130L} curve is not very different from the \textit{Herschel} extrapolation.
Compared to \citet{2019MNRAS.482.4906P}, this model predicts more \cii\ emitters at $L_{\rm \cii} > 10^{9.5} {\rm L}_{\odot}$, with luminosities consistent with the observed values for the clustered sample.
However, we see that our observed LFs (especially the clustered one) show a higher number density of objects ($> 1$ dex), which is not predicted by this model.
\subsection{Fitting with a Schechter function}
\label{sub:schech}
\begin{figure}
\centering
\includegraphics[scale=0.35]{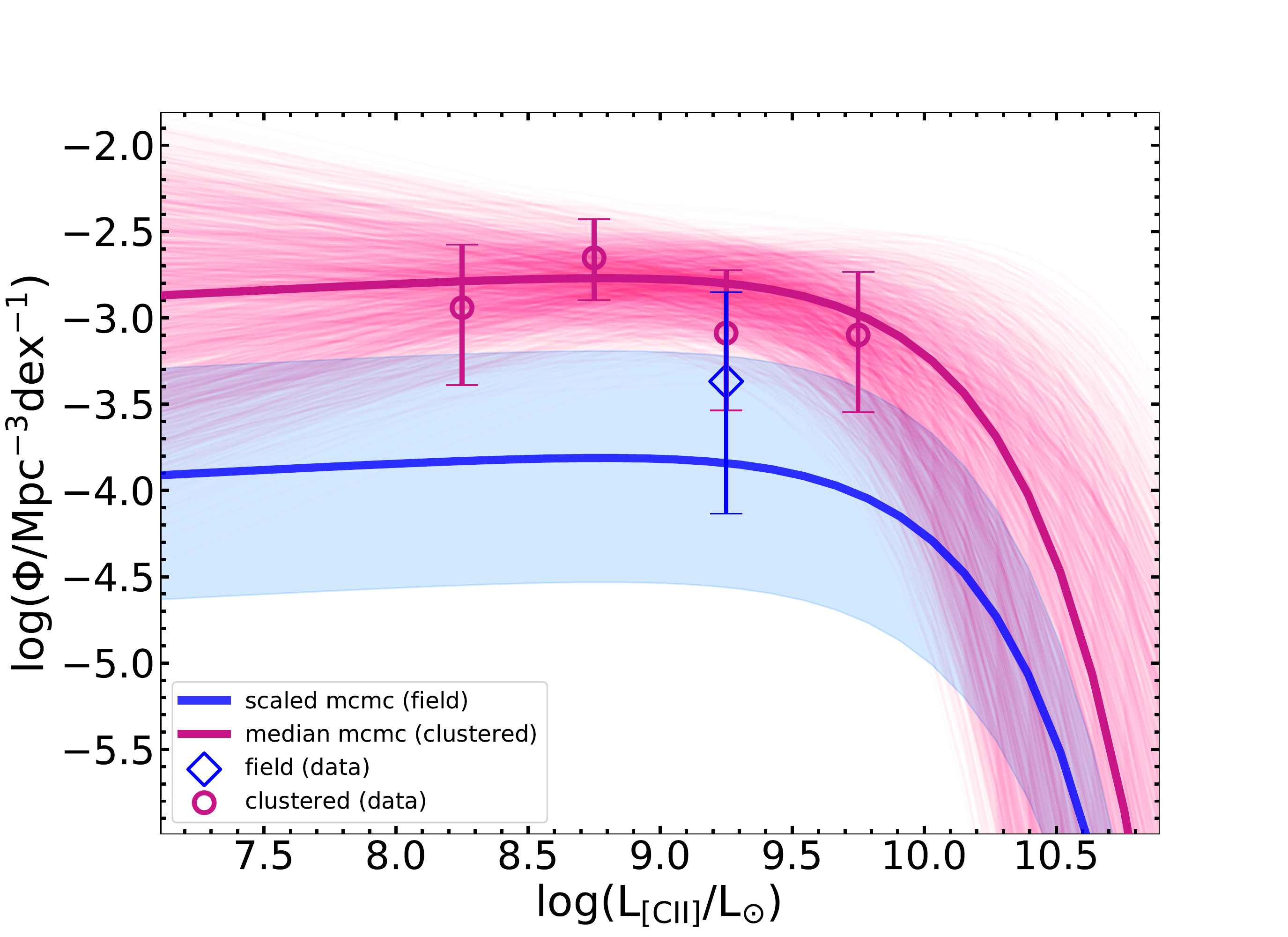}
\caption{Schechter functions for the clustered (pink) and field (azure) \cii\ luminosity functions (LFs). Also the observed LFs corresponding to the independent luminosity bins are indicated (same color code). We fitted $\log{\Phi^*}$, $\log{L^*}$, and $\alpha$ for the clustered LF using a Markov Chain Monte Carlo (MCMC) method. 
We assumed for the field sample the same $\alpha$ and $\log{L^*}$ of the clustered LF and we scaled the normalization of the clustered LF by a factor of 11 (corresponding to the ratio between the number of clustered/field sources). 
The shaded area (pink; clustered sample) shows the MCMC realizations within the 16th and 84th percentile, hence it corresponds approximately to 1$\sigma$ errorbars. In case of the scaled field LF the 1$\sigma$ errors (blue area) were computed from the uncertainties of $\log{\Phi^*}$ of the clustered sample and the Poissonian uncertainty (at 1$\sigma$) on 11 counts.}  
\label{fig:fit}
\end{figure}
We performed a fit to the \cii\ LFs with the \citet{1976ApJ...203..297S} function written in logarithmic form (Figure~\ref{fig:fit}). 
Given the element of luminosity d$\log{L}$, the number of objects $\Phi(L) {\rm d}\log{L}$ falling in the bin is:
\begin{equation}
\phi (L) {\rm d}\log{L} = \ln 10\ \Phi^* \left( \frac{L}{L^*} \right)^{\alpha + 1}\exp^{- \frac{L}{L^*}} \rm{d}\log{L}   
\end{equation}
where $\alpha$ is the faint--end slope and $L^*$ and $\Phi^*$ are the luminosity and the value of the LF at the $``$knee$"$ respectively.
For simplicity, we fitted the $\log{\Phi (L)}$ and thus also the logarithms of $L^*$ and $\Phi^*$.
We fitted the clustered LF only because of the low statistics of the field LF and the only one independent bin.
Before perfoming the fit, we rebinned the clustered and field LF adopting a bin spacing of 0.5 dex instead of 0.25 dex (see sec~\ref{sub:build}), i.e. equal to the bin width. This ensures that the number counts in the bins are independent as well as the uncertainties on the fitted points.\\
\begin{table}
\caption{Schechter parameters for the clustered  and field sample. We report the uncertainties corresponding to the 16th and 84th percentile ($\sim 1\sigma$).}             
\label{table:fit}      
\centering          
\resizebox*{0.35\textwidth}{!}{	
\begin{tabular}{c c c}     
\hline
\addlinespace[0.2cm]
\begin{small}Parameter\end{small}& \begin{small}Clustered\end{small} & \begin{small}Field\end{small} \\
& \begin{small}sample\end{small} & \begin{small} sample \end{small} \\
\hline
\addlinespace[0.2cm]
$\log{(L^*/\rm L_{\odot})}$ & $9.88^{+0.54}_{-0.55}$ & 9.88 (fixed) \\
\addlinespace[0.2cm]
$\log{(\Phi^*/\rm Mpc^{-3} \rm dex^{-1})} $ & $-3.01^{+0.44}_{-0.61}$ & $- 4.05^{+0.62}_{-0.72}$ \\
\addlinespace[0.2cm]
$\alpha$ & $-0.92^{+0.56}_{-0.44}$ & - 0.92 (fixed) \\
\addlinespace[0.2cm]
\hline       
\end{tabular}}
\end{table}
\begin{figure*}
\centering
\includegraphics[scale=0.6]{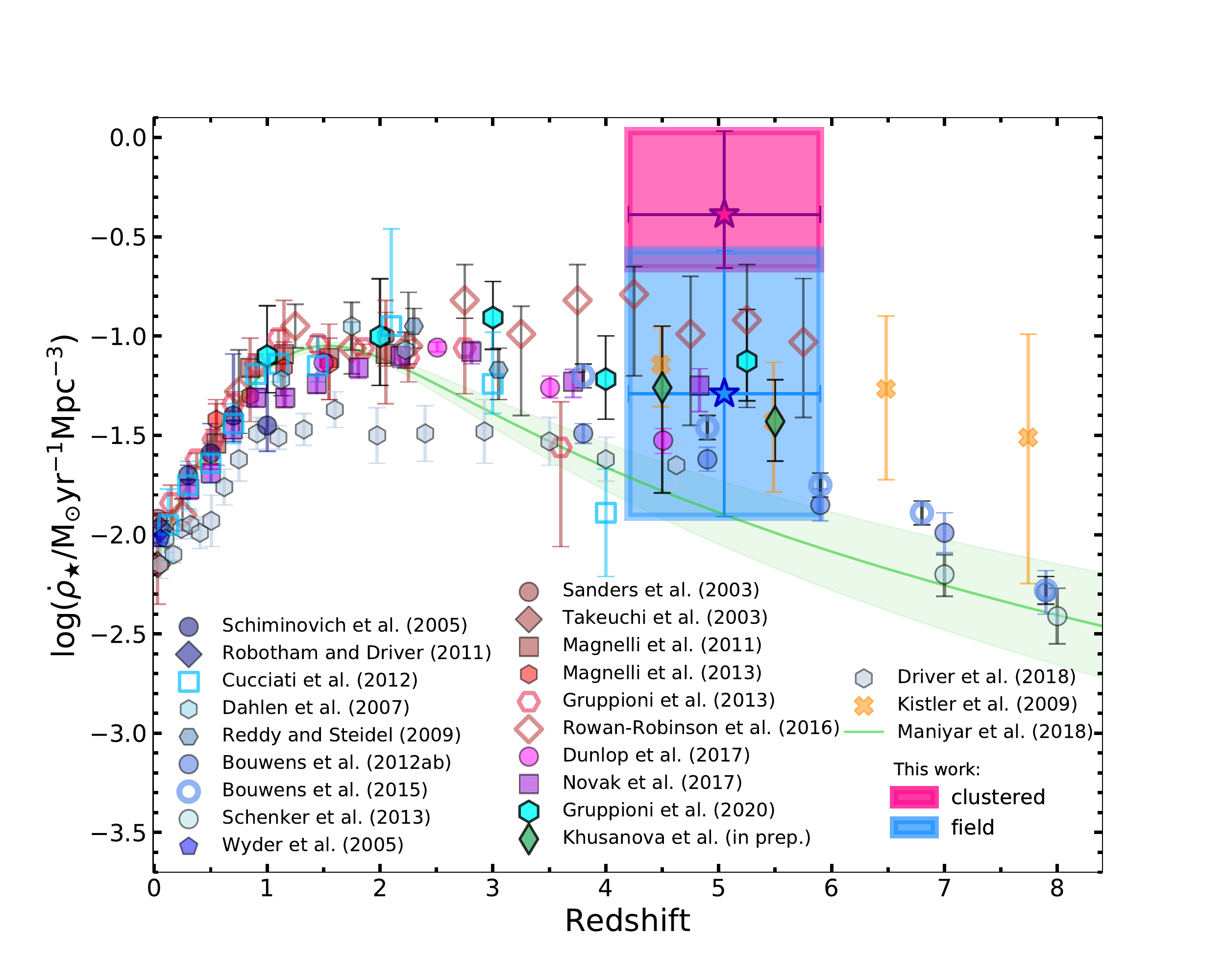}
\caption{Cosmic star formation rate density (SFRD) across cosmic time. Both the estimates from the $``$field$"$ and $``$clustered$"$ sample are shown (azure and pink box respectively).
We compare our measurements with estimates available from the literature based on multiwavenght observations. 
The SFRD derived from the clustered \cii\ LF at $z \sim 5$ is about one order of magnitude higher than the current measurements at that redshift. On the other hand, the SFRD of the field sample 
spans values compatible with both UV and IR--derived estimates, with an average value a factor $\sim 1.6$ higher than the estimates based on UV--surveys.
We consider the SFRD from the field sample as representative of the overall galaxy population since the clustered estimate is biased by overdensities around the targeted \cii .}  
\label{fig:sfrd}
\end{figure*}
\indent To derive a first estimate of the fitted parameters, we performed a fit based on the maximum likelihood criterion.
The best--fit values were used as initial guesses for a Markov Chain Monte Carlo (MCMC) method with the Python package \emph{emcee} \citep{2013PASP..125..306F}.
We assumed for $\alpha$, $\log{L^*}$, $\log{\Phi^*}$ uniform priors.
We preferred uniform than Gaussian priors as they represent the simplest possible choice, since the probability distribution of these parameters is not known a--priori. 
In case of $\log{L^*}$ we limited the upper boundary for the luminosity prior to 10.5, corresponding to an IR luminosity of $10^{13.5} \rm L_{\odot}$, assuming a fiducial ratio between \cii\ and IR luminosity of 10$^{-3}$ \citep{2013ApJ...774...68D}. This is a reasonable upper boundary to the IR luminosity motivated by pre--existing IR LFs at lower redshifts \citep{2013MNRAS.432...23G, 2016MNRAS.456L..40V}. The validity of the $L_{\rm \cii}$--SFR relation, where the latter quantity is derived from continuum estimates, for our sample (see Sect.~\ref{sub:cont}) suggests that this is a trustworthy assumption.\\
\indent The best values for $\alpha$, $\log{L^*}$, $\log{\Phi^*}$ for the clustered LF are reported in Table~\ref{table:fit}.
These values were evaluated as the medians of the posterior probability distributions. 
The reported uncertainties correspond to the 16th and 84th percentile of the posteriors (equivalent to about 1$\sigma$ in case of Gaussian posteriors).\\ 
\indent We then computed the Schechter function also for the field LF. Since it was not possible to directly fit the data, we scaled  $\Phi^*$ by a factor 1/11 (i.e. the ratio between the number of field and clustered sources), under the assumption that the shape of the two LFs is similar.
In this way the integration of the luminosity function over the accessible volume and luminosity predicts a number of sources equal to the observed one (i.e., one source). See also \citet{1983ApJ...269...42M}.
Moreover, this approach has the advantage to be independent from the binning of the LF. 
We obtained a value of $\log{(\Phi^*/\rm Mpc^{-3} \rm dex^{-1})} = - 4.05^{+0.62}_{-0.72}$ where the errors were propagated from the uncertainty on $\log{\Phi^*}$ of the clustered sample and the Poissonian error on the ratio 11:1. We note that the normalization determined in this way results consistent with the normalization that would be obtained by performing a Schechter fit to the field LF in which $\alpha$ and $\log{L^*}$ are fixed to the clustered values.\\ 
\indent The Schechter functions of the clustered and field samples were used to estimate the cosmic star formation rate density (Sect.~\ref{sec:sfrd}).
\section{Star formation rate density at $z\sim 5$}
\label{sec:sfrd}
We know that the \cii\ line is a SFR indicator \citep{2014A&A...568A..62D}. Therefore the \cii\ LF, providing the total \cii\ luminosity budget, can be used to estimate the cosmic star formation density (SFRD). First, we integrated the Schechter functions for the field and clustered sample in order to obtain the \cii\ luminosity density $\rho_{L_{\rm \cii}} = \int \Phi (L') L' \rm{d}\log{L'} $. We considered in the integration all the luminosities higher than $10^7 {\rm L}_{\odot}$.
However, integrating from lower luminosities does not alter significantly the final estimates because the LFs are quite flat.
In case of the clustered sample, the integration was performed for all the realizations of the MCMC. 
On the other hand, for the field sample, we integrated the best curve and the curves corresponding to the 1$\sigma$ errorbars. 
Then, we converted the luminosity densities to SFRDs using the relation (see Table 3 of \citealt{2014A&A...568A..62D}; case HII/starburst)
\begin{equation}
\log{\dot{\rho}_{\bigstar}} = - 7.06 + 1.00 \log{\rho_{L_{\rm \cii}}} + \log{0.94},
\end{equation}
where $\dot{\rho}_{\bigstar}$ is the SFRD and the last term accounts for scaling the \citet{2014A&A...568A..62D} relation from \citet{2001MNRAS.322..231K} to a \citet{2003PASP..115..763C} IMF.
We note that the working assumption of a non--evolving $L_{\rm \cii}$--SFR relation is not trivial \citep{2015ApJ...813...36V, 2018MNRAS.478.1170C}.
However, we mentioned in Sect.~\ref{sub:cont} that it seems to work at least for the serendipitous \cii\ detected in continuum. Furthermore, the validity of this conversion is independently confirmed by the ALPINE targets which, as discussed in \citet{2020arXiv200200962B} and \citet{2020arXiv200200979S}, lie within 1$\sigma$ on the \citet{2014A&A...568A..62D} relation.
In this way, we obtained for the clustered sample a SFRD probability distribution based on all the MCMC realizations. 
We considered the median value of the distribution as the best estimate of the SFRD from the clustered sample while, as done before, we reported the uncertainties corresponding to the 16th and 84th percentile (Table~\ref{table:sfrd}). On the other hand, for the field sample, we considered the SFRD value corresponding to the integration of the best curve with the associated errorbars (see Figure~\ref{fig:fit}). \\ 
\indent In Figure~\ref{fig:sfrd} we compare our results with previous estimates from the literature\footnote{For the works before 2014 we show the values reported in Table 1 of \citet{2014ARA&A..52..415M}, except for \citet{2009ApJ...705L.104K} that is not included in the table.}, based on UV surveys \citep{2005ApJ...619L..47S, 2005ApJ...619L..15W, 2007ApJ...654..172D, 2009ApJ...692..778R, 2011MNRAS.413.2570R, 2012ApJ...754...83B, 2012ApJ...752L...5B, 2012A&A...539A..31C, 2013ApJ...768..196S, 2015ApJ...803...34B} and IR, mm and radio selections of galaxies (\citealt{2003AJ....126.1607S, 2003ApJ...587L..89T, 2011A&A...528A..35M, 2013A&A...553A.132M, 2013MNRAS.432...23G, 2016MNRAS.461.1100R, 2017MNRAS.466..861D, 2017A&A...602A...5N}). We show also the measurements derived from optical--NIR observations \citep{2018MNRAS.475.2891D} and gamma--ray bursts \citep{2009ApJ...705L.104K}. 
We plot also the models predictions of \citet{2018A&A...614A..39M} based on the cosmic microwave background.
Finally, we compare our results with other independent measurements of the SFRD based on the ALPINE data. In particular, we show the results derived from the serendipitous sources detected in continuum (Gruppioni et al. 2020) and the SFRD inferred from the ALPINE central targets (Khusanova et al., in prep.).\\
\indent We can see that the SFRD derived from the clustered sample is almost $10\times$ higher than the field value and the current estimates of the SFRD at $z\sim 5$ from the literature.
We consider the SFRD computed using the field sample as the most likely estimate of the cosmic star formation activity at $z \sim 5$.
The measurement based on the clustered LF could be indeed biased by companions around the targeted \cii , which might not have been observed if we had started from a pure $``$blind$"$ survey. Therefore, the clustered estimate may not be representative of the overall population of galaxies.\\
\indent We know that a relevant question deals with the relative contribution of the unobscured wrt obscured star formation across cosmic time. The former is well sampled by UV surveys from $z\sim0$ up to $z\sim 10$ \citep{2015ApJ...803...34B, 2018ApJ...855..105O}. On the other hand, the latter is captured by surveys at longer wavelengths, typically IR and sub--mm. At the moment the obscured star formation is well constrained by statistically robust samples up to $z \sim 3$ while at higher redshift its contribution to the total budget of star formation is quite uncertain. 
If we look at the average value of the SFRD based on the field LF we can see that it is a factor $\sim 1.6$ higher than the measurement based on UV surveys \citep{2015ApJ...803...34B}. This means that it might be a fraction of (obscured) star formation that is not captured by UV surveys. However, when looking at the errors, we see that our estimate varies between values that are completely consistent with the UV estimates (i.e., neglible obscured star formation) to values that are about ten times higher than the UV measurements. A scenario consisting in a significant fraction of dust obscured star formation already in place at $z > 4$ is suggested by IR, mm and radio selections of galaxies (\citealt{2015ApJ...803...34B, 2017A&A...602A...5N}, Gruppioni et al. 2020).
Because of the large uncertainties our measurement does not allow us to assess the importance of obscured wrt unobscured star formation at $z\sim 5$.
Further observations of larger volumes in the sky are thus necessary to better constrain the \cii --derived SFRD.
\begin{table}
\caption{Cosmic star formation rate density (SFRD) from the clustered and field \cii\ LFs.}             
\label{table:sfrd}      
\centering          
\resizebox*{0.35\textwidth}{!}{	
\begin{tabular}{c c c }     
\hline
\addlinespace[0.2cm]
& \begin{small}Clustered\end{small} & \begin{small}Field\end{small} \\
\hline
\addlinespace[0.2cm]
 $\log{(\frac{\dot{\rho}_{\bigstar}}{\rm M_{\odot} \rm yr^{-1} \rm Mpc^{-3}})}$ & $-0.39^{+0.42}_{-0.27}$ & $-1.29^{+0.72}_{-0.62}$ \\
\addlinespace[0.2cm]
\hline       
\end{tabular}}
\end{table}
\section{Summary and conclusions}
\label{sec:con}
We summarize the main results of this work:

\begin{enumerate}

\item We built the \cii\ luminosity function (LF) using the lines serendipitously discovered in the ALMA ALPINE large program.
This is the first LF at $z \sim 5$ based on galaxies purely selected based on their \cii\ line emission.
First of all, we performed a blind search in the 118 ALPINE pointings that revealed several unexpected lines.
We assessed the fidelity and the completeness of the detections. 
The final catalog of the serendipitous sources includes 14 line emitters with high fidelity ($> 85$\% for 12 out of 14 detections). 
We identified the line emission by comparing its spatial position with the available photometric catalogs and multi--wavelength images.
Out of the 14 lines, eight are \cii\ lines at  $4.3 < z < 5.4$, supported by a spectroscopic or photometric redshift from ancillary data. Two out of 14 lines are CO transitions at lower redshift. Finally four out of 14 lines have a more tricky nature because they are not associated to any optical/NIR counterpart or have an uncertain photometric redshift. However, three of them are very likely \cii\ emitters based on the strict association with the central target or individual SED modelling. Observational follow-up are necessary to unambiguosly confirm the nature of these sources.\\ 

\item The eight \cii\ emitters and the four lines with an ambiguos identification were used to build the \cii\ LF. 
We found that 11 out of 12 sources are strongly clustered around the central target in the same poining since they are located at very similar redshifts ($|\Delta z| < 0.0154$, corresponding to $|\Delta v| < 750$ \kms ). 
The discovery of these sources could be very useful to investigate the properties of overdense regions at high--$z$ and their study will be exhaustively addressed in a future work. 
In order to take the clustering into account when building the \cii\ LF, we split our sample in two (i.e. a $``$clustered$"$ and $``$field$"$ sub--samples) based on the redshift separation between the serendipituous line and the central target in the same pointing and built two separate LFs.
The median luminosity of the field and clustered samples is very similar; however, the clustered LF shows luminosities a factor $\sim 10$ higher than the field one. 
If this trend were confirmed by a larger sample of galaxies it could highlight the already known dependence between clustering and luminosity, witnessed for the first time from the \cii\ line at $z\sim 5$.\\

\item We compared our LFs with other works, both observational and theoretical. 
We found that, globally, the clustered LF suggests an excess of sources compared to the LFs of the ALPINE targets (Yan et al. 2020). This could be due both to clustering effect and to the fact that the targets LFs are based on UV--selected galaxies, hence they do not include highly dusty objects.
On the other hand the field LF is quite consistent with the targets LFs.
Our measurements, especially the field one, are quite in agreement with the estimates from the serendipitous continuum sources found in ALPINE (Gruppioni et al. 2020). 
The estimates from the field LF are also in agreement with the semi--analytical models of \citet{2019MNRAS.482.4906P} at $L_{\rm \cii} \sim 10^{9} {\rm L}_{\odot}$. Observations of more extended volumes will be useful to assess if this agreement persists also at higher luminosities.
Finally, both the clustered and field LFs suggest a possible evolution of the \cii\ LF from $z\sim 5$ to $z\sim 0$. Also in this case observations of larger samples are necessary to confirm this trend.\\

\item We performed a Schechter fit to the clustered LF. 
Then, we scaled the fitted normalization function for a factor 11 (corresponding to the ratio between the number of clustered and field sources) to reproduce the field LF, under the assumption that they have the same shape. From the Schechter fits we estimated the \cii\ luminosity density. This value was then converted to star formation rate density (SFRD) using the relation of De Looze et al. (2014) (case HII/starburst). We found that the clustered sample shows values that are up to a factor $\sim 10$ higher than the current estimates from the literature. We considered the estimate obtained for the field sample as representative of the average galaxy population, since the clustered estimate could be biased to high--density environment. The average SFRD results a factor $\sim 1.6$ on average higher than the current estimates from UV surveys. However, because of the large errorbars, it is not possible to say if this value could be indicative of significant fraction of obscured star formation at $z \sim 5$, as suggested by IR and mm selections of galaxies. Observations of larger samples are necessary to better constrain the SFRD of the Universe.

\end{enumerate}

This work represents the first effort to study the luminosity function of \cii\ emitters at $z\sim 5$ based on sources purely selected due to their \cii\ emission and provides new insight onto the properties of the star--forming medium in a poorly known cosmic epoch.

\begin{acknowledgements}
This paper is based on data obtained with the ALMA Observatory, under Large Program 2017.1.00428.L. ALMA is a partnership of ESO (representing its member states), NSF (USA) and NINS (Japan), together with NRC (Canada), MOST and ASIAA (Taiwan), and KASI (Republic of Korea), in cooperation with the Republic of Chile. The Joint ALMA Observatory is operated by ESO, AUI/NRAO and NAOJ. Based on data products from observations made with ESO Telescopes at the La Silla Paranal Observatory under ESO programme ID 179.A-2005 and on data products produced by TERAPIX and the Cambridge Astronomy Survey Unit on behalf of the UltraVISTA consortium.
This program is supported by the national program Cosmology and Galaxies from the CNRS in France. 
F.L., C.G., A.C., F.P. and M.T. acknowledge the support from the grant PRIN MIUR 2017.
D.R. acknowledges support from the National Science Foundation under
grant numbers AST-1614213 and AST-1910107 and from the Alexander von
Humboldt Foundation through a Humboldt Research Fellowship for
Experienced Researchers.
G.C.J. acknowledges ERC Advanced Grant 695671 ``QUENCH'' and support by the Science and Technology Facilities Council (STFC). 
GL acknowledges support from the European Research Council (ERC) under the European Union’s Horizon 2020 research and innovation programme (project CONCERTO, grant agreement No 788212) and from the Excellence Initiative of Aix-Marseille University-A*Midex, a French “Investissements d’Avenir” programme.
\end{acknowledgements}

%
%
\bibliographystyle{aa}
\bibliography{bibliografia} 

\begin{appendix} 
\section{Images and spectra}
\label{app:spec}
\begin{figure*}
\centering
\includegraphics[scale=0.78]{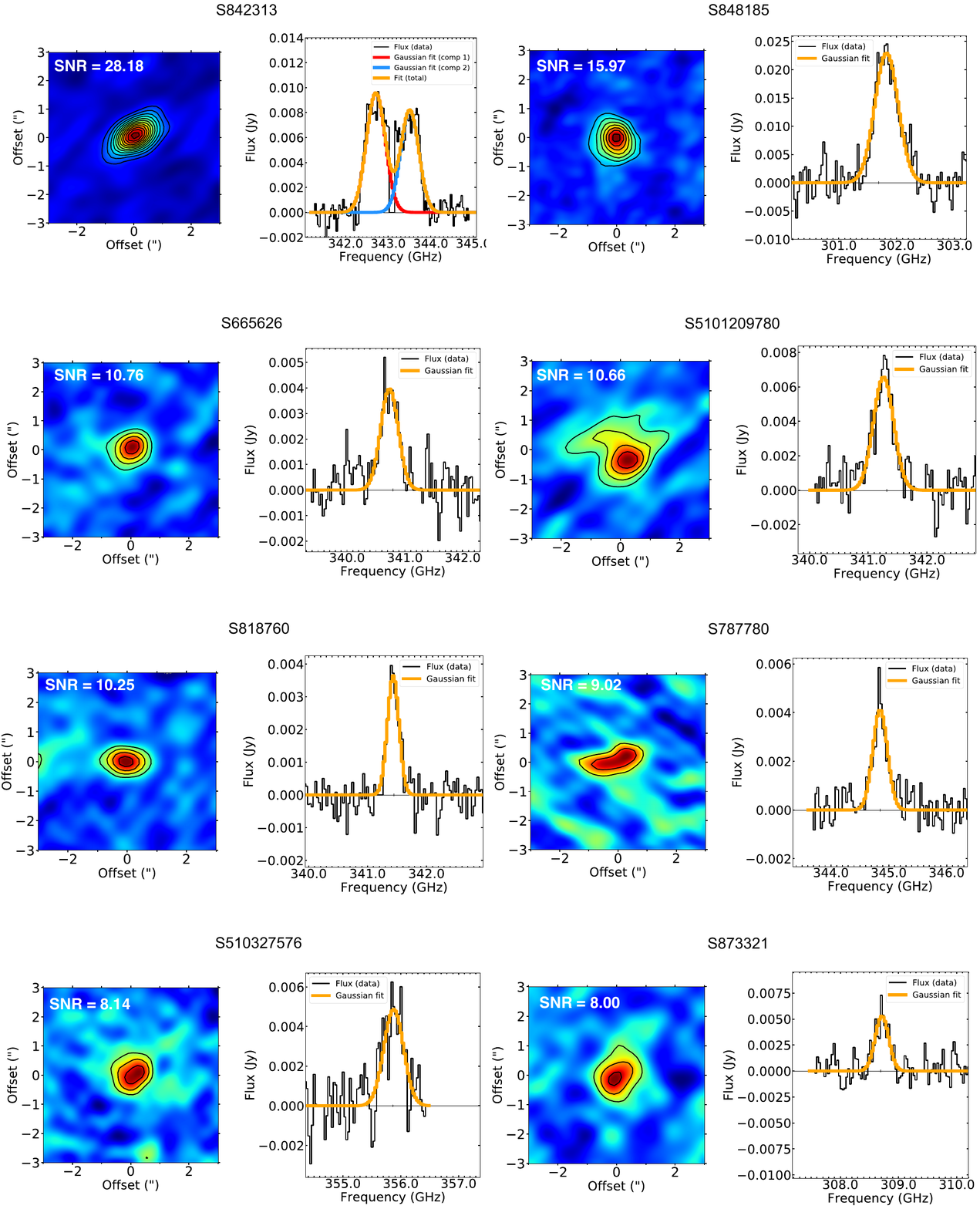}
\caption{Continumm--subtracted ALMA maps and spectra of the 14 serendipitous lines found in the 118 ALPINE pointings. Each panel is labelled according to the number of the ALPINE source in the same pointing. The lowest contour level corresponds to 3$\sigma$. We fitted the line emission (black) using a sigle Gaussian component (orange). In case of S842313 we fitted the line profile using two Gaussian components (cyan and red); the total model is shown in orange. For S5100822662 the serendipitous source is the small blob above the ALPINE. The blob is marked with a cross and the spectrum shows its emission.}
\label{fig:spectra}
\end{figure*}
\begin{figure*}
\centering
\ContinuedFloat
\includegraphics[scale=0.78]{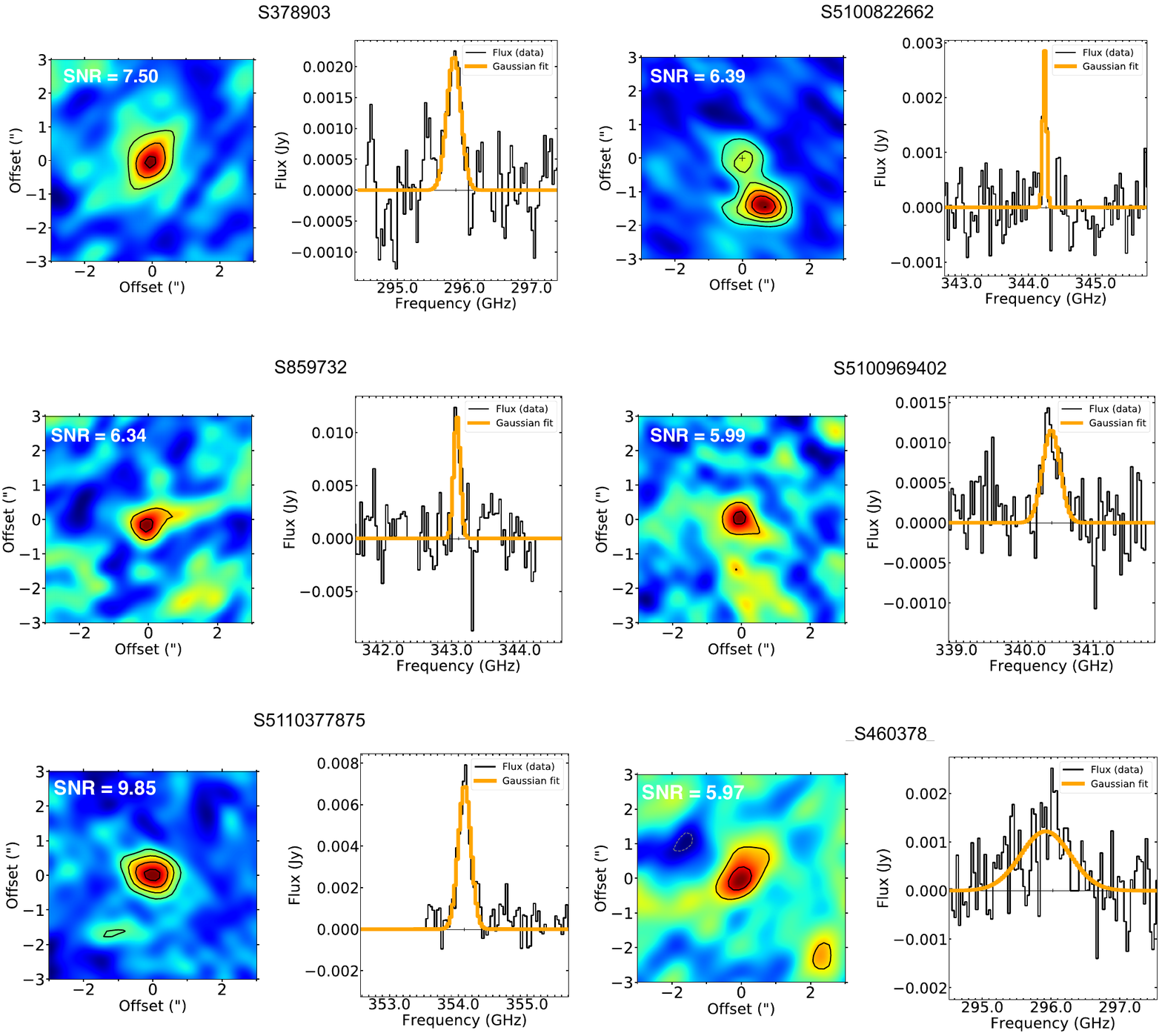}
\caption{(continued) Continumm--subtracted ALMA maps and spectra of the 14 serendipitous lines found in the 118 ALPINE pointings.}
\label{fig:spectra}
\end{figure*}
\section{Completeness}
\label{app:comp}
We show the completeness curves as a function of the flux peak for fixed FWHM in Appendix~\ref{app:comp}. We can see that at fixed flux peak the completeness is higher for larger lines. 
\begin{figure*}
\centering
\includegraphics[scale=0.3]{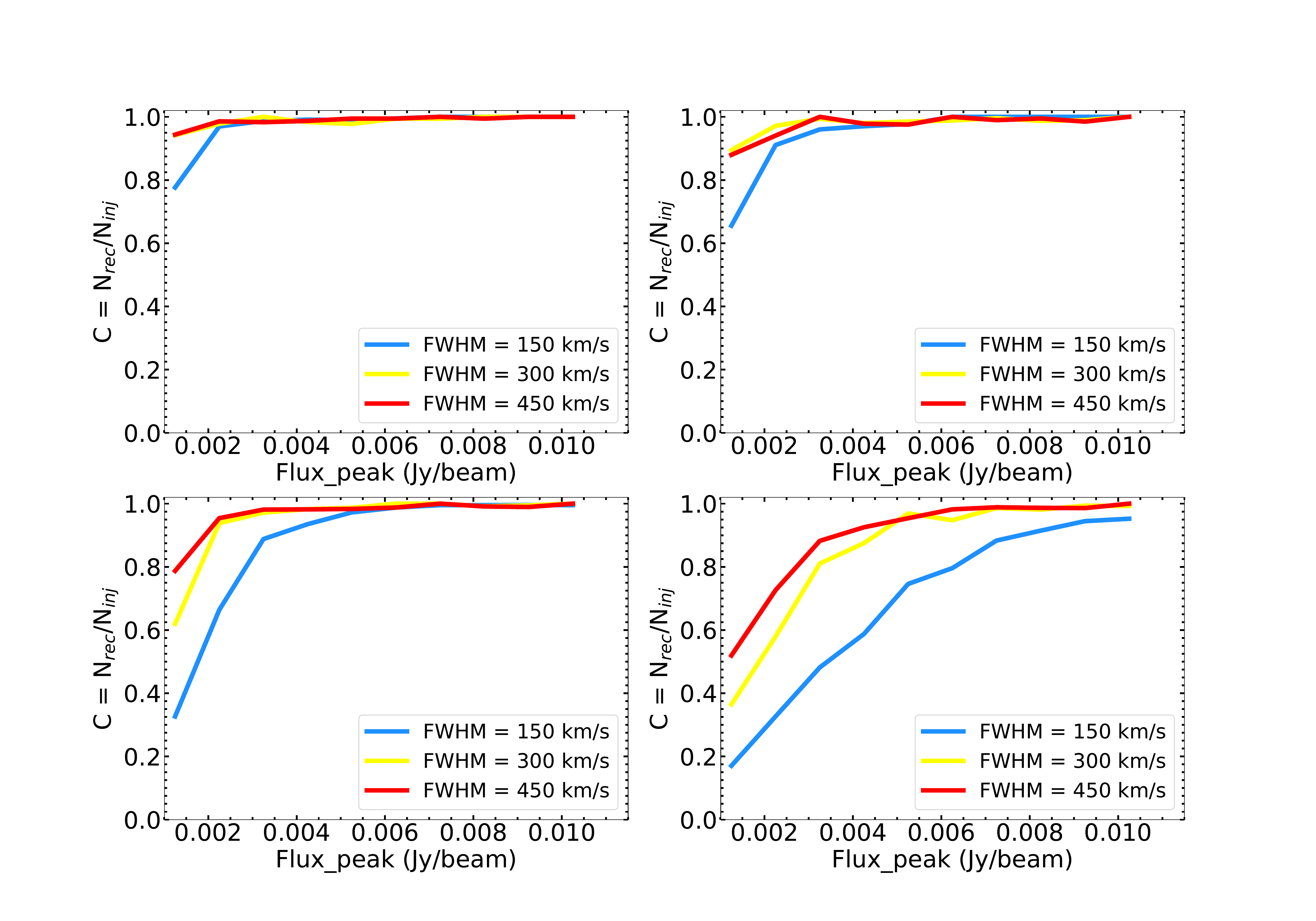}
\caption{Completeness curves for fixed FWHM in four regions with radii $R_{<30}$ (distance from the phase center $R_{<30} \leq 7.1 "$), $R_{30-50}$ ($ 7.1 " < R_{30-50} \leq 10.4 "$), $R_{50-70}$ ($10.4 " < R_{50-70} \leq 13.1 "$), $R_{70-90}$ ($ 13.1 " < R_{70-90} \leq 16.4 "$), defined on the basis of their primary beam attenuation. The curves correspond to 150, 300, 450 \kms (reported in blue, yellow and red respectively). We can see that completeness is a strong function of the line location and width.}
\label{fig:complCurv}
\end{figure*}
\section{Luminosity function}
\label{app:lf}
\begin{figure}
\centering
\includegraphics[scale=0.38]{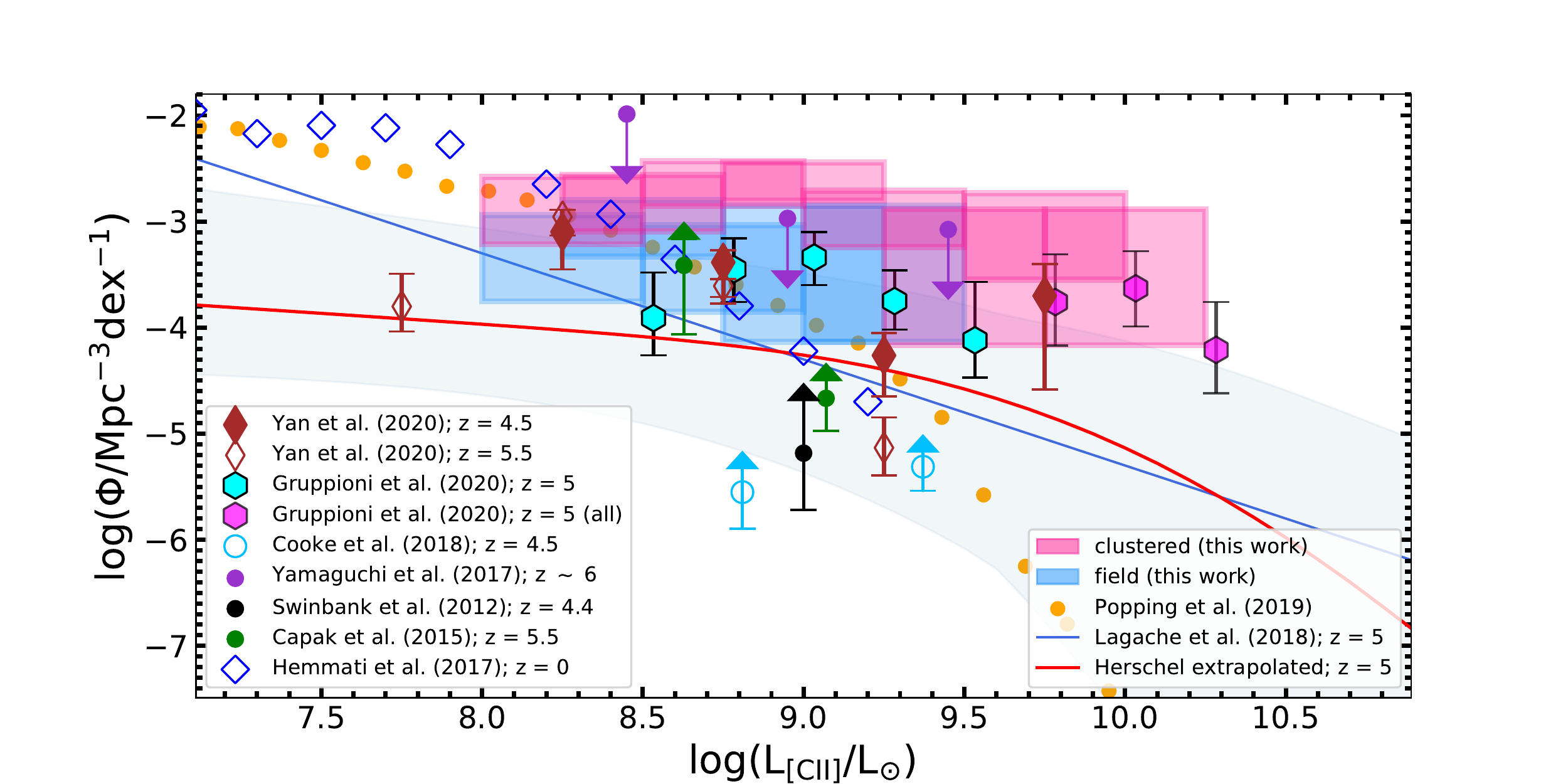}
\caption{\cii\ luminosity functions (LFs) at $z \sim 5$ of the serendipitous sources in ALPINE including the sources with low fidelity. Compared to the clustered LF, the field one lies at lower luminosity. Both LFs result compatible with those based on the sources with higher fidelity. The main difference is at the low luminosity end since more sources are included in this sample.}
\label{fig:lf_dook}
\end{figure}

\begin{figure}
\centering
\includegraphics[scale=0.38]{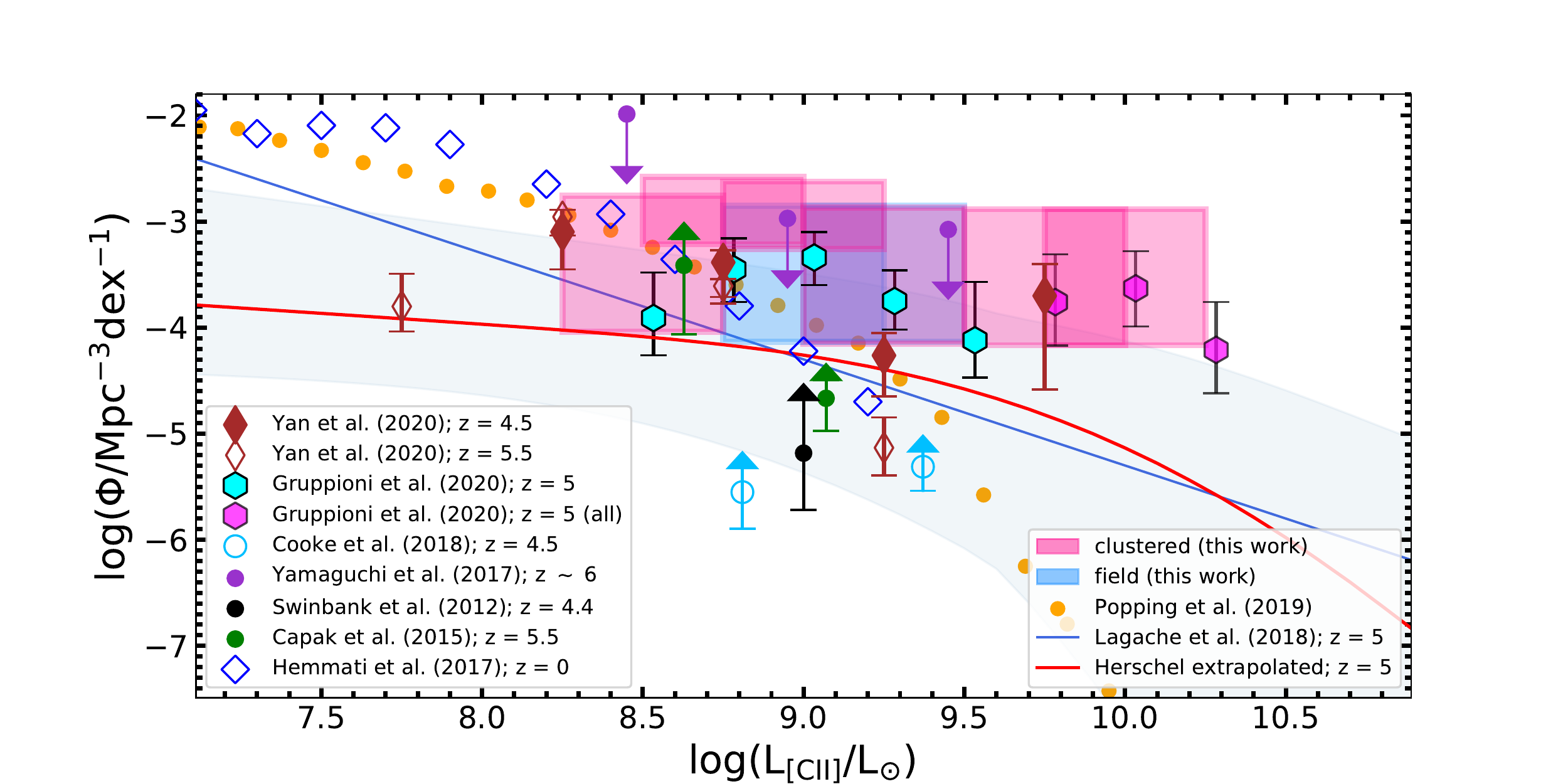}
\caption{\cii\ luminosity function (LF) at $z \sim 5$ of the clustered sources, after removing the six \cii --emitters associated to interacting galaxies with the central targets (see Figure~\ref{fig:cutouts}). Compared to the LF presented in Sect.~\ref{sec:lf}, this one results lower at the faint end.}
\label{fig:lf_large}
\end{figure}

We show in this section the \cii\ luminosity function (LF) computed including sources with a fidelity $>$20\% (Figure~\ref{fig:lf_dook} and Table~\ref{table:lf_dook}). 
In Sect.~\ref{sub:fidelity} we included in our catalog only the detections with fidelity $> 85$\%, with only one \cii\ emitter (S5100969402) having a fidelity $\sim 50$\%, in order to study a very robust sample. However this sample is obviously incomplete at the lowest luminosities. 
We thus derived the clustered and field LFs for two new sub--samples, in which we included also low fidelity (i.e. low luminosity) sources. The fidelity cut of 20\% adds 10 lines to our catalog of serendipitous detection. Based on their individual fidelities we expect that only $\sim 3$ out of the 9 lines are genuine detections, thus this new sample is expected to contain a high fraction of spurious sources.
One line is possibly associated to CO emission at $z < 5$ (see Sect.~\ref{sec:ide}). The remaining nine do not present any optical/NIR counterpart hence their redshift is unconstrained from ancillary data. Therefore, for these sources we can only assume that their emission is due to the \cii\ line. 
We see that the shape of both clustered/field LFs remains quite unchanged, with the field LF sampled by more sources now. The field sources lie at lower luminosities compared to the clustered sample. However, this is not surprising since, as the fidelity lowers, also the lines flux decreases and hence we expect the population of the low luminosity bins only for both field and clustered sources.\\ 
\indent We show also the \cii\ LF (clustered sample) obatined excluding the six \cii\ possibly associated to mergers/interactions (Figure~\ref{fig:lf_large} and Table~\ref{table:lf_large}). 
The new points are consistent within the errors with the old ones.
\begin{table}
\caption{Luminosity functions for the clustered and field sample considering the 21 sources with fidelity higher than 20\%. We reported also the number of sources in each luminosity bin. We indicated with the bold font the values corresponding to the independent bins.}             
\label{table:lf_dook}      
\centering          
\resizebox*{0.45\textwidth}{!}{	
\begin{tabular}{c c c c c}     
\hline
\addlinespace[0.2cm]
$\log{(L/\rm L_{\odot})}$&$\log{(\Phi_{\rm clust}/\rm Mpc^{-3} \rm dex^{-1})}$& $N_{\rm clust}$ &$\log{(\Phi_{\rm field}/ \rm Mpc^{-3} \rm dex^{-1})}$ & $N_{\rm field}$ \\
\addlinespace[0.2cm]
\hline
\addlinespace[0.2cm]
 8.25& $\textbf{-2.87}^{\textbf{+0.29}}_{\textbf{-0.34}}$ & \textbf{3} & $\textbf{-3.30}^{\textbf{+0.36}}_{\textbf{-0.45}}$ & \textbf{2} \\
\addlinespace[0.2cm]
8.50 & $-2.81^{+0.25}_{-0.28}$ & 4 & $-3.05^{+0.25}_{-0.28}$ & 4 \\
\addlinespace[0.2cm]
8.75& \textbf{$\textbf{-2.64}^{\textbf{+0.20}}_{\textbf{-0.22}}$} & \textbf{6} & $\textbf{-3.40}^{\textbf{+0.36}}_{\textbf{-0.45}}$ & \textbf{2} \\
\addlinespace[0.2cm]
9.00& $-2.62^{+0.17}_{-0.18}$ & 8 & $-3.37^{+0.52}_{-0.77}$ & 1 \\
\addlinespace[0.2cm]
9.25& \textbf{$\textbf{-2.96}^{\textbf{+0.25}}_{\textbf{-0.28}}$} & \textbf{4} & \textbf{$\textbf{-3.37}^{\textbf{+0.52}}_{\textbf{-0.77}}$} & \textbf{1} \\
\addlinespace[0.2cm]
9.50& $-3.40^{+0.52}_{-0.77}$ & 1 &  ... & ... \\
\addlinespace[0.2cm]
9.75& \textbf{$\textbf{-3.10}^{\textbf{+0.36}}_{\textbf{-0.45}}$} & \textbf{2} & ... & ... \\
\addlinespace[0.2cm]
10.00& $-3.40^{+0.52}_{-0.77}$ & 1 & ... & ... \\
\addlinespace[0.2cm]
\hline       
\end{tabular}}
\end{table}

\begin{table}
\caption{Luminosity functions for the clustered sample after having excluded the six sources interacting with the central targets. We reported also the number of sources in each luminosity bin. We indicated with the bold font the values corresponding to the independent bins.}             
\label{table:lf_large}      
\centering          
\resizebox*{0.3\textwidth}{!}{	
\begin{tabular}{c c c}     
\hline
\addlinespace[0.2cm]
$\log{(L/\rm L_{\odot})}$&$\log{(\Phi_{\rm clust}/\rm Mpc^{-3} \rm dex^{-1})}$& $N_{\rm clust}$ \\
\addlinespace[0.2cm]
\hline
\addlinespace[0.2cm]
8.25&...&...\\
\addlinespace[0.2cm]
8.50 & $-3.27^{+0.52}_{-0.77}$ & 1  \\
\addlinespace[0.2cm]
8.75& \textbf{$\textbf{-2.88}^{\textbf{+0.29}}_{\textbf{-0.34}}$} & \textbf{3}  \\
\addlinespace[0.2cm]
9.00& $-2.92^{+0.29}_{-0.34}$ & 3  \\
\addlinespace[0.2cm]
9.25& \textbf{$\textbf{-3.39}^{\textbf{+0.52}}_{\textbf{-0.77}}$} & \textbf{1} \\
\addlinespace[0.2cm]
9.50& ... & ... \\
\addlinespace[0.2cm]
9.75& $\textbf{-3.40}^{\textbf{+0.52}}_{\textbf{-0.77}}$ & \textbf{1} \\
\addlinespace[0.2cm]
10.00& $-3.40^{+0.52}_{-0.77}$ & 1 \\
\addlinespace[0.2cm]
\hline       
\end{tabular}}
\end{table}


\end{appendix}
%
%
\end{document}